\documentclass[showpacs,preprintnumbers,nofootinbib]{revtex4}
\usepackage{amsmath,amsfonts,amssymb,amsbsy,bbm,graphicx,epsfig,rotating}
\usepackage[dvips]{color}
\newcommand{\blue}{\textcolor{blue}}
\newcommand{\red}{\textcolor{red}}
\newcommand{\green}{\textcolor{green}}

\newcommand{\be}{\begin{equation}}
\newcommand{\ba}{\begin{eqnarray}}
\newcommand{\ee}{\end{equation}}
\newcommand{\ea}{\end{eqnarray}}
\newcommand{\mn}{\mu\nu}
\newcommand{\half}{\frac{1}{2}}

\newcommand{\barr}{\begin{array}}
\newcommand{\ear}{\end{array}}
\newcommand{\vphi}{\phi}

\newcommand{\tT}{\frac{t-t_0}{T}}

\newcommand{\dphi}{\delta\phi}
\newcommand{\dbeta}{\delta\beta}
\newcommand{\dsigma}{\delta\sigma}

\newcommand{\dnu}{\delta\nu}
\newcommand{\dz}{\delta z}

\newcommand{\bea}{\begin{array}}
\newcommand{\eea}{\end{array}}
\def\bal{{\mbox{\boldmath $\alpha$}}}
\def\bq{{\bf q}}
\def\bd{{\bf d}}
\def\bp{{\bf p}}
\def\d{\delta}
\def\e{\epsilon}

\begin{document}

\title{Cosmological Solutions of Low-Energy Heterotic M-Theory\\}

\author{Edmund J. Copeland}
\email{ed.copeland@nottingham.ac.uk} \affiliation{School of Physics and Astronomy,
University of Nottingham, University Park, Nottingham, NG7 2RD, UK}
\author{James Ellison}
\email{j.b.ellison@sussex.ac.uk} \affiliation{Department of Physics and Astronomy,
University of Sussex, Brighton, BN1 9QJ, UK}
\author{Andre Lukas}
\email{lukas@physics.ox.ac.uk} \affiliation{Rudolf Peierls Centre for Theoretical Physics,
University of Oxford, Oxford OX1 3NP, UK}
\author{Jonathan Roberts}
\email{j.roberts@sussex.ac.uk} \affiliation{Department of Physics and Astronomy, University
of Sussex, Brighton, BN1 9QJ, UK}

\date{\today}

\begin{abstract}
\vskip 0.3cm

\noindent We derive a set of exact cosmological solutions to the $D=4$, $\mathcal{N}=1$
supergravity description of heterotic M-theory. Having identified a new and exact $SU(3)$
Toda model solution, we then apply symmetry transformations to both this solution and to a
previously known $SU(2)$ Toda model, in order to derive two further sets of new cosmological
solutions. In the symmetry-transformed $SU(3)$ Toda case we find an unusual ``bouncing''
motion for the M5 brane, such that this brane can be made to reverse direction part way
through its evolution. This bounce occurs purely through the interaction of non-standard
kinetic terms, as there are no explicit potentials in the action. We also present a
perturbation calculation which demonstrates that, in a simple static limit, heterotic
M-theory possesses a scale-invariant isocurvature mode. This mode persists in certain
asymptotic limits of all the solutions we have derived, including the bouncing solution.
\end{abstract}


\maketitle

%===============================================INTRODUCTION==========================================
%=====================================================================================================

\section{Introduction}

In the past ten years, heterotic M-theory has provided an exciting arena in which to analyse
the cosmology and particle physics of our
universe~\cite{Horava:1996ma,Bandos:1997ui,Lukas:1998tt,Brandle:2001ts,Khoury:2001wf,Steinhardt:2002ih}.
Representing the low-energy limit of the strongly-coupled heterotic $E_{8} \times E_{8}$
string, this theory not only combines gravitational, particle and braneworld physics into
one unified description, but also possesses a detailed and constrained field-content that
cannot be arbitrarily adjusted. Therefore, it retains a definitive and unambiguous
relationship to M-theory itself.  However, in the $D=4, \mathcal{N}=1$ supergravity
description of heterotic M-theory, a number of important questions still remain unanswered.
Consider, for example, the simple cosmological situation that occurs when we retain only the
dilaton $S$, the universal $T$ modulus, and the field $Z$ describing a single M5 brane.
Despite the fact that this leads to vanishing superpotential in four dimensions, the
resulting cosmology is highly non-linear and demonstrates quite unexpected behaviour. In
Ref.~\cite{Copeland:2001zp} the $S,T,Z$ cosmology was analysed in a truncated limit with all
axion fields removed. It was then shown that the scalar corresponding to the M5 brane
position \emph{must} be included in the set of cosmologically significant fields, and this
leads to a forcing effect whereby the ambient dimensions change size as the brane moves.
Moreover, the frictional forces acting back on the brane are such that it accelerates and
then decelerates back to rest, mimicking a time-dependent force of finite duration. This
illustrates the unconventional effect of non-standard kinetic terms in the theory, and the
means by which the brane can undergo a single displacement by exchanging energy with its
environment. In the special situation presented in Ref.~\cite{Copeland:2001zp} this effect
can be described exactly using the Toda formalism \cite{Kostant:1979qu,Lukas:1996iq}, and
the model of Ref.~\cite{Copeland:2001zp} is an $SU(2)$ Toda model.

Given this behaviour, it is interesting to consider whether more complicated trajectories
for the M5 brane are possible.  For example, not all of the scalar fields were considered in
the $SU(2)$ Toda model of Ref.~\cite{Copeland:2001zp}, since the axionic fields were
consistently truncated away. This means that only a portion of the full solution-space was
explored, and that the $SU(2)$ behaviour is liable to be only an approximation once the
axions are restored. Therefore, following on from the work of Ref.~\cite{Copeland:2001zp},
we wish to analyse situations in which additional axionic fields are evolving in conjunction
with the brane, and determine whether interesting new behaviours for the M5 brane can occur.
In particular, we wish to determine whether an M5 brane can undergo multiple displacements,
and even reverse direction in the absence of explicit potentials.

Before embarking on the detailed calculations, we first summarise our results. We uncover a
new and exact $SU(3)$ Toda model, in which the M5 brane can undergo two successive
displacements in the same direction. That is, the brane spontaneously accelerates twice in
response to the other moduli fields to which it is coupled. Applying the symmetries derived
in our companion paper Ref.~\cite{Copeland:2005mk} to this model, as well as to the known
$SU(2)$ model of Ref.~\cite{Copeland:2001zp}, we obtain two additional sets of new
solutions. In the symmetry-transformed $SU(3)$ case the brane can undergo two successive
displacements in \emph{opposite} directions, and so reverse direction and ``bounce'' without
the presence of any explicit potentials in the action. This effect occurs purely through the
interaction of non-standard kinetic terms, via the cross-couplings of the various fields,
and constitutes an exact supergravity solution that has been rigorously deduced from
M-theory. Finally we investigate the generation of density perturbations in these models,
and show that heterotic M-theory possesses a scale-invariant isocurvature mode in some of
the axion fields. This last result is consistent with the original findings of the pre Big
Bang scenario \cite{Copeland:1997ug,Copeland:1998ie} and in agreement with the result
obtained in Ref.~\cite{DiMarco:2002eb} where it was first shown that the moving brane itself
could not generate a scale-invariant perturbation spectrum. Following a conclusion, in an
appendix we present the technical details of the $SU(3)$ Toda model derivation.

%============================================THEORY==========================================
%=============================================================================================

\section{The Four-dimensional action}\label{sec:4daction}

We now review the $D=4, \mathcal{N}=1$ supergravity action presented in
Ref.~\cite{Copeland:2001zp}. Recall that this was derived via a compactification of 11D
supergravity on the orbifold $S^1/\mathbb{Z}_{2} \times CY_{3}$, where $CY_{3}$ denotes a
Calabi-Yau three-fold. This leads to two four-dimensional boundary planes separated along a
fifth dimension. If the fifth dimension is labelled by a normalised coordinate $z \in
[0,1]$, then the boundaries reside at $z=0,1$ respectively and have the charges
$q_{0},q_{1}$. A single M5 brane is also included in the space, by wrapping it on a
holomorphic 2-cycle of the $CY_{3}$. The brane then appears as a three-brane of charge $q$
that lies parallel to the boundaries, and which can move along the interval. Importantly,
the interaction between the boundaries and brane leads to the existence of a static,
triple-domain wall BPS solution. One can then consider further reducing on this solution, so
as to find a supergravity theory describing slowly varying fluctuations about the static BPS
vacuum. This contains the six scalar fields $\beta,\chi,\phi,\sigma,z,\nu$ with the
following non-standard kinetic terms
\begin{align}
\begin{split}
S_{4}=-\frac{1}{2\kappa_{4}^2}\int_{M_{4}} d^{4}x\sqrt{-g}\left[
\frac{1}{2}R+\frac{3}{4}(\partial\beta)^{2}+3e^{-2\beta}(\partial\chi)^{2}+\frac{1}{4}(\partial\phi)^{2}+
\frac{1}{4}e^{-2\phi}\left(\partial\sigma + 4qz\partial\nu\right)^{2}
 \right.\\
\left.+\frac{1}{2}qe^{\beta-\phi}(\partial z)^{2}+2qe^{-\beta-\phi}(\partial\nu-\chi
\partial z)^{2} \right]\label{4daction}
\end{split}
\end{align}
Each of these scalars has an underlying significance in terms of the $D=5$ parent theory
from which it descends. The scalar $\beta$ is the zero-mode of the $g_{55}$ component in the
$D=5$ metric, and measures the separation between the boundaries. Specifically, the
separation is given by $\pi\rho e^{\beta}$ in terms of some dimensionful reference size
$\pi\rho$. The field $\phi$ represents the orbifold-averaged Calabi-Yau volume, such that
the physical size is given by $v e^{\phi}$ in terms of a dimensionful reference volume $v$.
The scalars $\sigma,\chi$ originate from the bulk three-form and graviphoton field
respectively. The field $z$ measures the position of the bulk brane between the boundaries,
with the points $z=0,1$ corresponding to the boundaries. Lastly, the field $\nu$ arises from
the self-dual two-form on the brane worldvolume.

This reduction on a BPS solution guarantees that the scalars must group into supersymmetric
multiplets described by a supersymmetric action. One can verify that they naturally fall
into the pairs $(\phi,\sigma)$,$(\beta,\chi)$,$(z,\nu)$, which are the bosonic components of
chiral superfields $S,T,Z$ as follows
\begin{align}\label{complexstructure}
S= e^{\phi} +qz^{2}e^{\beta} + i\left(\sigma +2qz^2\chi\right) \quad ,\quad T= e^{\beta}
+2i\chi \quad,\quad Z= e^{\beta}z + 2i(-\nu+ z\chi)
\end{align}
This naturally leads to a K\"ahler manifold expression for the scalar part of the action
\begin{align}
S_{4}=-\frac{1}{2\kappa_{4}^{2}}\int d^{4}x \sqrt{-g} \left(\frac{1}{2}R+
K_{i\bar{j}}\:\partial_{\mu}{\Phi}^{i} \partial^{\mu}{\bar{\Phi}}^{\bar{j}}\right)
\end{align}
where the superfields are grouped into a coordinate vector $\Phi=(S,T,Z)$, with the complex
conjugate coordinates denoted by $\bar{\Phi}$. The K\"ahler metric $K_{i\bar{j}}$ is given
by
\begin{align}
K_{i\bar{j}} = \frac{ \partial^{2}{K} }{\partial{\Phi^{i}}\partial{\bar{\Phi}^{\bar{j}}}}
\end{align}
in terms of the K\"ahler potential
\begin{align}\label{kahlerpot}
K = -\ln\left[S+\overline{S}-q\frac{(Z+\overline{Z})^{2}}{T+\overline{T}}\right]
-3\ln\left(T+\overline{T}\right)
\end{align}
This K\"ahler potential is computed only to linear-order in the two parameters
$\epsilon_{k}$ $(k=1,2)$ defined by
\begin{align*}
\epsilon_k= \sum_{i=0}^{k-1} \: \pi
\left(\frac{\kappa}{4\pi}\right)^{2/3}\frac{2\pi\rho}{v^{2/3}}\:q_{i}\:e^{\beta-\phi}
\end{align*}
Here $\kappa$ is the eleven-dimensional Newton constant, and $\pi\rho,v$ are the
dimensionful scales mentioned above. The two conditions $\epsilon_{k} \ll1$ then restrict
the accessible regions of moduli-space in which we can trust the four-dimensional effective
theory. In addition, the supergravity action Eq.~\eqref{4daction} can only be trusted in the
limit where stringy $\alpha'$ corrections are suitably small, as these corrections introduce
higher-derivative terms that we have disregarded.

%=============================================EQUATIONS OF MOTION=====================================
%=====================================================================================================

\section{Equations of Motion}\label{sec:eqnsmotion}

We now turn to the equations of the motion arising from the action Eq.~\eqref{4daction}. If
we assume a spatially flat Friedmann Robertson Walker (FRW) cosmology for the
four-dimensional spacetime, then the metric takes the form
\begin{align}
ds_{4}^{2} = -e^{2n(\tau)}d\tau^{2} + e^{2\alpha(\tau)}\delta_{ij}dx^{i}dx^{j}
\end{align}
where $i,j=1,\dots,3$, the scale-factor is $\alpha(\tau)$, and $n(\tau)$ represents a gauge
freedom in the choice of time coordinate. Denoting a $\tau$ derivative by an overdot, and
assuming all fields are purely functions of $\tau$, one obtains the Einstein field equations
\begin{align}
-3\dot{\alpha}^2 + \frac{1}{4}\dot{\phi}^{2}+\frac{3}{4}\dot{\beta}^{2}
+\frac{1}{2}qe^{\beta-\phi}\dot{z}^{2} +
3e^{-2\beta}\dot{\chi}^{2}+&\frac{1}{4}e^{-2\phi}\left(\dot{\sigma} +
4qz\dot{\nu}\right)^{2} +2qe^{-\beta-\phi}(\dot{\nu}-\chi\dot{z})^{2}
=0 \\
2\ddot{\alpha}+(3\dot{\alpha}-2\dot{n})\dot{\alpha} +
\frac{1}{4}\dot{\phi}^{2}+\frac{3}{4}\dot{\beta}^{2} +\frac{1}{2}qe^{\beta-\phi}\dot{z}^{2}
+ 3e^{-2\beta}\dot{\chi}^{2}+&\frac{1}{4}e^{-2\phi}\left(\dot{\sigma} +
4qz\dot{\nu}\right)^{2} +2qe^{-\beta-\phi}(\dot{\nu}-\chi\dot{z})^{2}=0
\end{align}

the $\phi,\beta,z$ equations of motion
\begin{align}
\ddot{\phi}+(3\dot{\alpha}-\dot{n})\dot{\phi} + qe^{\beta-\phi}\dot{z}^{2} +
(\dot{\sigma}+4qz\dot{\nu})^{2}e^{-2\phi}+4q(\dot{\nu}-\chi\dot{z})^{2}e^{-\beta-\phi} &=0 \\
3\ddot{\beta}+ 3(3\dot{\alpha}-\dot{n})\dot{\beta} - qe^{\beta-\phi}\dot{z}^{2} +
12\dot{\chi}^2e^{-2\beta} + 4q(\dot{\nu}-\chi\dot{z})^{2}e^{-\beta-\phi}&=0\\
\frac{d}{d\tau}\left\{
\left[e^{\beta-\phi}\dot{z}-4\chi(\dot{\nu}-\chi\dot{z})e^{-\beta-\phi} \right]e^{3\alpha-n}
\right\}e^{n-3\alpha} -2\dot{\nu}(\dot{\sigma}+4qz\dot{\nu})e^{-2\phi} &=0
\end{align}

and the $\sigma,\nu,\chi$ equations of motion
\begin{align}
\frac{d}{d\tau}\left\{ \: \left[ \: (\dot{\sigma}+4qz\dot{\nu})e^{-2\phi} \: \right]e^{3\alpha-n} \: \right\} &= 0\\
\frac{d}{d\tau}\left\{ \left[ z(\dot{\sigma}+4qz\dot{\nu})e^{-2\phi} +
2(\dot{\nu}-\chi\dot{z})e^{-\beta-\phi} \right]e^{3\alpha-n} \right\} &= 0\\
\frac{d}{d\tau}\left[ \: (3e^{-2\beta}\dot{\chi})e^{3\alpha-n} \: \right]e^{n-3\alpha}
+2q\dot{z}(\dot{\nu}-\chi\dot{z})e^{-\beta-\phi}&=0
\end{align}

As we cannot exactly solve these equations of motion, we will utilise the following two
solution methods. Firstly, we search for specialised solutions that occur when the equations
are truncated, usually by setting certain combinations of fields to zero. This will allow us
to recover the known $SU(2)$ Toda model of Ref.~\cite{Copeland:2001zp}, as well as a
previously undiscovered $SU(3)$ Toda model. Secondly, we will utilise the scalar-field
symmetry transformations that were derived in our recent companion paper
Ref.~\cite{Copeland:2005mk}. That is, we will apply these symmetry transformations to the
fields of the $SU(2)$ and $SU(3)$ Toda models in turn, and so derive two new solutions to
the equations of motion. We will find that in these new solutions the M5 brane can evolve
in far more complicated ways than has previously been seen.

\section{Review of the $SU(2)$ Toda Model}\label{sec:su2}

We now briefly recall the behaviour of the $SU(2)$ Toda model found in
Ref.~\cite{Copeland:2001zp}. This will prove useful because the $SU(2)$ model exhibits
features that persist in all the solutions we will present, and so will illuminate the
discussions to come. In addition, it is worthwhile studying this model in order to
understand how symmetry-transformations will affect it.

\subsection{The $SU(2)$ Toda Model Solutions}

The $SU(2)$ model can be derived by choosing the axions to satisfy
$\sigma,\nu=\mathrm{constant},\chi=0$. The remaining fields $\alpha,\phi,\beta,z$ can then
be solved for exactly, essentially due to the fact that the field $z$ satisfies the
conservation law
\begin{align}\label{eq:zconserve}
e^{\beta-\phi+3\alpha-n}\dot{z} = \mathrm{constant}
\end{align}
Namely, inserting this result back into the remaining equations of motion yields a closed
set of equations in $\alpha,\beta,\phi$ which can be solved in isolation. In particular,
these equations can be reformulated in terms of the motion of a ``particle'' with
coordinates $\mbox{\boldmath$\alpha$}=(\alpha,\beta,\phi)$ which roams over a
three-dimensional space and experiences an exponential potential $U$. These equations take
the form
\begin{align}
\frac{d}{d\tau}\left(EG\mbox{\boldmath$\dot{\alpha}$}\right) + E^{-1}\frac{\partial
U}{\partial \mbox{\boldmath$\alpha$}}=0 \quad, \quad
\frac{1}{2}E\mbox{\boldmath$\dot{\alpha}$}^{T}G\:\mbox{\boldmath$\dot{\alpha}$} +E^{-1}U=0
\end{align}
which can be derived by variation of $\mbox{\boldmath$\alpha$},E$ from the simpler particle
Lagrangian
\begin{align}\label{eq:particlelag}
\mathcal{L}=\frac{1}{2}E\mbox{\boldmath$\dot{\alpha}$}^{T}G\:\mbox{\boldmath$\dot{\alpha}$}
-E^{-1}U
\end{align}
Here we have defined a moduli-space metric $G=\mathrm{diag}(-3,\frac{3}{4},\frac{1}{4})$ of
Minkowski signature, and a particle-worldline metric $E=e^{-n
+\mathbf{d}\cdot\mbox{\boldmath$\alpha$}}$ where \mbox{$\mathbf{d}=(3,0,0)$} is a dimension
vector that gives the number of spatial dimensions associated to the scale-factor $\alpha$.
The function $E$ thus encodes the arbitrary choice of time parameterisation of the
worldline, with its variation in the Lagrangian naturally producing an energy conservation
constraint. Finally, the moduli-space potential $U=U_{1}$ is given by
\begin{align}
U_{1} &=
\frac{1}{2}u_{1}^{2}\:\mathrm{exp}\left(\mathbf{q}_1\cdot\mbox{\boldmath$\alpha$}\right)
\quad,\quad \mathbf{q}_1=(0,-1,1)
\end{align}
where $u_{1}^{2}$ is a positive constant. Up to a constant length rescaling, the vector
$\mathbf{q}_1$ defines the single, simple root-vector of the Lie algebra $SU(2)$, and so the
Lagrangian Eq.~\eqref{eq:particlelag} defines an $SU(2)$ Toda model. This is an exactly
integrable system, and in the proper-time gauge $n=0$ one finds the general solution:
\begin{align}\label{eq:su2solns}
 \bal-\bal_0 = {\bf p}_i \ln \left| \frac{t-t_0}{T} \right| + \left( {\bf p}_f -{\bf
p}_i \right) \ln \left( 1 + \left| \frac{t-t_0}{T} \right|^{- \delta} \right)^{-1/\delta}
\end{align}
\begin{align} \label{z}
z-z_0 =  d \left( 1 + \left| \frac{t-t_{0}}{T} \right|^{\delta} \right)^{-1}
\end{align}
\begin{align}~\label{eq:su2cons}
 \bp_{\gamma} G\bp_{\gamma} = 0 \quad, \quad \bp_{\gamma} \cdot \bd = 1 \quad , \quad
 \bq_1 \cdot \bal_0 = \ln\left(\frac{q d^2 <\bq_1, \bq_1>}{8} \right) \quad, \quad \d = -\bq_1
 \cdot\bp_i\quad, \quad \bp_f - \bp_i = \delta \frac{2G^{-1}\bq_1}{<\bq_1 ,\bq_1 >}
\end{align}
Here the subscript $\gamma$ takes the values $\gamma=i,f$, and the scalar product $<\cdot
,\cdot>$ is defined by $<{\bf a},{\bf b}> = {\bf a}^T G^{-1} {\bf b}$. Note that the
constraints in Eq.~\eqref{eq:su2cons} must be enforced so that
Eqs.~\eqref{eq:su2solns}-\eqref{z} are indeed the correct field solutions.  Once this is
done, the solution describes a transition between two asymptotically free-field states. That
is, the initial field velocities are equal to the ``expansion power'' constants ${\bf
p}_i=(1/3,p_{\beta,i},p_{\phi,i})$, the final field velocities are equal to the constants
${\bf p}_f=(1/3,p_{\beta,f},p_{\phi,f})$, and the nonsupersymmetric second term in
Eq.~\eqref{eq:su2solns} forces a smooth acceleration between these ``rolling-radii'' (rr)
regimes. In fact, the underlying reason for this ${\bf p}_i \rightarrow {\bf p}_f $
interpolation is the motion of the brane itself, which according to Eq.~\eqref{z} is at rest
in the extreme limits, but borrows kinetic energy from $\beta,\phi$ and moves significantly
at the intermediate time $t-t_0\approx T$. For the sake of concreteness, we now present the
explicit form of the solutions by inserting the various vector quantities. This gives
\begin{align*}
\alpha-\alpha_0&=\frac{1}{3}\ln \left| \frac{t-t_{0}}{T} \right|\\
\beta-\beta_0&= p_{\beta,i}\ln \left| \frac{t-t_{0}}{T} \right| + (p_{\beta,f}-p_{\beta,i})
\ln \left( 1 + \left| \frac{t-t_0}{T} \right|^{- \delta} \right)^{-1/\delta}\\
\phi-\phi_0 &= p_{\phi,i}\ln \left| \frac{t-t_{0}}{T} \right| + (p_{\phi,f}-p_{\phi,i})\ln
\left( 1 + \left| \frac{t-t_0}{T} \right|^{- \delta} \right)^{-1/\delta}
\end{align*}
These are subject to the relations
\begin{align*}
\delta=p_{\beta,i}-p_{\phi,i} \quad, \quad %
\beta_{0}-\phi_{0} = \ln\left(\frac{3}{2qd^2}\right)\quad ,\quad %
 \left(\begin{array}{c}
         p_{\beta ,f}\\
         p_{\phi ,f}
         \end{array}\right) =
 \frac{1}{2}\left( \begin{array}{rr}
                        1&1\\
                        3&-1
                    \end{array}\right)
 \left(\begin{array}{c}
       p_{\beta ,i}\\
       p_{\phi ,i}
       \end{array}\right)
\end{align*}
as well as the  ``ellipse'' constraint
\begin{align}\label{eq:ellipse}
3p_{\beta,i}^2+p_{\phi,i}^2=\frac{4}{3}
\end{align}
Notice, in fact, that if we enforce the constraint that fixes $(p_{\beta,f},p_{\phi,f})$ in
terms of $(p_{\beta,i},p_{\phi,i})$ then the final expansion powers
$(p_{\beta,f},p_{\phi,f})$ are automatically guaranteed to lie on the same ellipse. The
interpretation of this ellipse condition is relatively simple, and can be illustrated in a
phase-plane diagram as follows. Consider drawing a 2D plot where the horizontal axis
corresponds to $d\beta/d u$ (with $u \equiv 3\alpha$) and the vertical axis to $d\phi/d u$.
Then the constants $(p_{\beta,\gamma},p_{\phi,\gamma})$, where $\gamma=i,f$, are the
asymptotic values of $d\beta/d u$ and $d\phi/du$. Thus, they correspond to the values
attained in the plane at the extreme endpoints of the phase-plane trajectory. Hence,
according to Eq.~\eqref{eq:ellipse} and the constraints, the trajectory traced out in the
phase-plane must begin and end at two different points on a single, fixed ellipse drawn in
that plane. These phase-plane plots, or ``ellipse diagrams'' as we shall call them, will
prove to be extremely useful in exhibiting the behaviour of the system diagrammatically.
This is because the shape and curves of the trajectories in these diagrams tell us very
graphically about the brane motion and changes to the axions.

\subsection{Analysis and Validity of the $SU(2)$ Toda Model}

To exhibit the behaviour of the fields, we now plot an ellipse diagram. This proves to be
far more intuitive and useful than following the behaviour of all fields individually.
Before doing this, we must recognise that the solutions in Eq.~\eqref{eq:su2solns} are valid
over two disconnected time ranges given by
\begin{align}
 t \in \left\{ \begin{array}{clll}
                    \left( -\infty ,t_0\right)\; , &(-) \; {\rm branch} \\
                    \left( t_0,+\infty\right)\; ,&(+)\;{\rm branch}
               \end{array} \right.
\end{align}
where the time $t=t_0$ corresponds to a curvature singularity. Consequently, there are two
different notions of ``early'' and ``late''  built into the solutions, depending on the
choice of branch. For example, although $t=t_{0}$ corresponds to a \emph{past} singularity
in the $(+)$ branch, it corresponds to a \emph{future} singularity from the perspective of
the $(-)$ branch. The $(-)$ branch is, in fact, an example of a pre Big Bang (PBB) era,
which is automatically undergoing superluminal deflation.

\begin{figure}[htbp]
\begin{center}
\includegraphics[width=6cm]{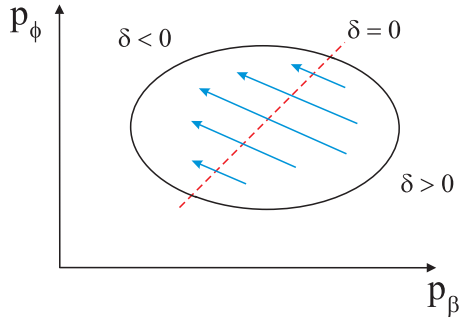}
\caption[$SU(2)$ ellipse and field behaviour]{\label{fig:su2ellipse} \textit{Figure to show
the linear mapping across the ellipse, with the direction fixed by the choice of time
branch. If we pick a candidate set of expansion powers satisfying $\delta>0$, then these are
indeed available early-time states on the $(-)$ branch. States satisfying $\delta<0$ are not
available at early time; instead, they are the late time states that the system evolves
into. This is all reversed on the $(+)$ branch.}}
\end{center}
\end{figure}

Therefore, to avoid confusion we must always pick a particular branch, and take care with
what constitutes early and late behaviour. In particular, the $\mathbf{p}_i$ constants only
correspond to an ``initial'' set of expansion powers as implied by the subscript if they
satisfy
\begin{align}
\delta = p_{\beta,i}-p_{\phi,i} \left\{ \begin{array}{c}
                                                        > 0  \quad (-) \: {\rm branch} \\
                                                        < 0  \quad (+) \: {\rm branch}
                                                        \end{array} \right.
\end{align}
That is, only those powers satisfying this condition can ever be early time states of the
system. In Fig.~\ref{fig:su2ellipse} we have plotted some representative trajectories on the
$(-)$ branch. Note that the fields $\beta,\phi$ start at a single point on the ellipse, with
their initial powers corresponding to that sector with $\delta>0$. On the negative branch
this ``early'' time state corresponds to the infinitely negative past $t-t_0 \rightarrow
-\infty$. The fields then evolve such that the effective trajectory in the plane is a
straight line. This linear behaviour is a consequence of the relation $3d\beta/du+d\phi/du
\propto \mathrm{constant}$, and this in turn is possible because all the axions have been
truncated away. The trajectory then ends on the opposite sector of the ellipse, ending up in
a ``late'' time state as $t-t_0 \rightarrow 0$ from below. This directed evolution between
parts of the ellipse cannot be reversed unless we switch the branch from $(-)$ to $(+)$, so
the accessible early-time states of the system are fixed by the choice of the branch. Thus,
interesting physical results will sometimes necessitate choosing one branch over another.

To complete this section, we now comment on the validity of these $SU(2)$ Toda solutions.
Recall that the four-dimensional action Eq.~\eqref{4daction} is known only as a power
series, with five-dimensional gravitational corrections measured in powers of the
$\epsilon_{k}$ ($k=1,2$). In the above model one finds that
\begin{align}
 \epsilon_{k} \sim \left\{ \begin{array}{lll }
                          |t-t_0|^{+\delta} &\rightarrow \infty \quad \text{early time}\\
                          |t-t_0|^{-\delta} &\rightarrow \infty \quad \text{late time}
                          \end{array}\right.
\end{align}
This divergence is due to the fact that the coupling of the bulk brane to $\beta,\phi$ is
itself proportional to the $\epsilon_{k}$, so that when the brane moves the system is
necessarily driven to a five-dimensional limit in both asymptotic regimes. This will cause
the four-dimensional theory to break down, and with it the solutions
Eq.~\eqref{eq:su2solns}. This divergence is in fact familiar from PBB cosmology where the
addition of the dilatonic axion to the dilaton-moduli system causes the same problem. A more
insidious problem, however, is that even at intermediate times one cannot make $\epsilon_{k}
\ll1$ whilst simultaneously fitting the entire $z$ displacement profile within the physical
orbifold extent $z\in[0,1]$. Say, for example, that we search for the minimum value of the
$\epsilon_{k}$ parameters. One can verify that this occurs at precisely the time
$t-t_{0}=T$, and at this time the ellipse trajectory intersects the line $\delta=0$. The
magnitude at the minimum is then
\begin{align*}
\epsilon_{k}|_{\mathrm{min}} \sim \frac{1}{d^{2}}
\end{align*}

Evidently, $d\gg1$ is now required in order to make this small. Having made this choice,
there will be a finite period of time, depending on the value of $d$, where the $\epsilon_k$
are small and our solutions, Eq.~\eqref{eq:su2solns}, are valid. This period ends with the
rapid collision of the brane with the boundary which, of course, also invalidates the above
analytical solutions. (For a discussion of the evolution after the collision see
Ref.~\cite{Gray:2003vk}). The limitations on the validity of these solutions, both due to
the $\epsilon_k$ constraint and brane-boundary collision, may be viewed as a disadvantage
and one may ask whether other solutions with a larger range of validity exist. We will show
that this is indeed the case for some of the new solutions to be discussed below.

%===============================================SU3 TODA MODEL======================================================

\section{The $SU(3)$ Toda Model}

After reviewing the $SU(2)$ model at some length, we will now present an entirely new
$SU(3)$ solution. This proves to be significantly more complicated than the $SU(2)$ Toda
model, as we might expect from the fact that the Lie group $SU(3)$ is more complicated in
structure than $SU(2)$. We will find that the $SU(3)$ solutions are fundamentally controlled
by two influences, one due to the motion of the M5 brane and the other due to changes in
$\chi$. This leads to new, characteristic trajectories in the ellipse diagrams. In
particular, the coupling between $z$ and $\chi$ allows the M5 brane to undergo two
successive displacements in the same direction. This generalises the behaviour of the old
$SU(2)$ case, and demonstrates that the brane can undergo repetitious displacement
behaviour.

\subsection{The $SU(3)$ Toda Model Solutions}
To derive the $SU(3)$ Toda model, one chooses the axions to satisfy
\begin{align}\label{eq:todatrunc}
\dot{\nu}-\chi\dot{z}= \dot{\sigma}+4qz\dot{\nu} = 0
\end{align}
This effectively sets two of the terms in the action to zero for all time, without forcing
any of the fields or their derivatives to be \emph{individually} zero. To see that this
corresponds to an $SU(3)$ Toda model we impose Eq.~\eqref{eq:todatrunc}, and note that the
$z$ and $\chi$ equations are now total derivatives
\begin{align*}
\frac{d}{d\tau}\left[e^{\beta-\phi+3\alpha-n}\dot{z}\right]=
\frac{d}{d\tau}\left[e^{-2\beta+3\alpha-n}\dot{\chi}\right]=0
\end{align*}
These can be immediately integrated to give constants of the motion. Inserting these
conservation laws back into the remaining equations of motion then yields a closed set of
equations in $\alpha,\beta,\phi$ that can again be derived from the particle Lagrangian
\begin{align}~\label{eq:su3partlagrangian}
\mathcal{L}=\frac{1}{2}E\mbox{\boldmath$\dot{\alpha}$}^{T}G\:\mbox{\boldmath$\dot{\alpha}$}-E^{-1}U
\end{align}
The quantities $\mbox{\boldmath$\alpha$},G,E$ remain unchanged from the $SU(2)$ Toda model,
but the potential is modified to $U=U_{1}+U_{2}$, with
\begin{align*}
U_{1} &=
\frac{1}{2}u_{1}^{2}\:\mathrm{exp}\left(\mathbf{q}_{1}\cdot\mbox{\boldmath$\alpha$}\right)
\quad,\quad \mathbf{q}_{1}=(0,-1,1) \\
U_{2} &=
\frac{1}{2}u_{2}^{2}\:\mathrm{exp}\left(\mathbf{q}_{2}\cdot\mbox{\boldmath$\alpha$}\right)
\quad,\quad \mathbf{q}_{2}=(0,2,0)
\end{align*}
This means that the effective particle motion of \mbox{\boldmath $\alpha$} is now subjected
to \emph{two} exponential forces. To be a Toda model, a precise relationship must exist
between the orientations and lengths of the vectors defined by the $\mathbf{q}_{i}$.
Consider the following matrix
\begin{align}\label{eq:su3cartan}
<\mathbf{q}_{i},\mathbf{q}_{j}> = \frac{8}{3}\cdot \left( \begin{array}{cc}
                                                    2 & -1 \\
                                                    -1 & 2 \\
                                                    \end{array} \right)
\end{align}
where the scalar product is once again defined by $<{\bf a},{\bf b}> = {\bf a}^T G^{-1} {\bf
b}$. The right hand side of Eq.~\eqref{eq:su3cartan} happens to be a constant multiple of
the Cartan matrix of the $SU(3)$ Lie algebra, so that up to a constant length rescaling the
vectors $\mathbf{q}_{1},\mathbf{q}_{2}$ are identical to the two simple root-vectors of
$SU(3)$. Thus, the model is an exactly integrable $SU(3)$ Toda model. In particular, an
exact, analytical description of the behaviour is now accessible if we choose a basis for
the moduli-space that is adapted to the $SU(3)$ root-vectors. This decouples the equations
of motion and allows them to be readily solved, the complicated details of which are
reserved for Appendix A. The proper-time solutions for the fields in the gauge $n=0$ are
then given by

\begin{align}
\bal-\bal_0 = {\bf p}_i \ln\left| \frac{t-t_0}{T} \right| & + \left( {\bf p}_f -{\bf p}_i
\right) \ln \left[ 1 + \left| \frac{t-t_0}{T} \right|^{-\delta}\left(1+
\theta_z^2 \left|\frac{t-t_{0}}{T_{\beta}}\right|^{-\delta_{\beta}}\right)\right]^{-1/\delta} \nonumber\\
+&\left( {\bf p}_{f}^{(\chi)} -{\bf p}_{i} \right) %
\ln \left[1 + \left|\frac{t-t_0}{T_{\beta}} \right|^{-\delta_{\beta}}\left(1+
\theta_{\chi}^2 \left|\frac{t-t_{0}}{T}\right|^{-\delta}\right)\right]^{-1/\delta_{\beta}}
\end{align}
\begin{align}
z-z_0&=d \left( 1+ \left|\frac{t-t_0}{T_{}}\right|^{\delta} +
\theta_{z}^2\left|\frac{t-t_0}{T_{\beta}}\right|^{-\delta_{\beta}} \right)^{-1}\cdot
\left( 1+\theta_{z}\left|\frac{t-t_0}{T_{\beta}}\right|^{-\delta_{\beta}}\right)\\
\chi-\chi_0 &= d_{\chi} \left( 1+ \left|\frac{t-t_0}{T_{\beta}}\right|^{\delta_{\beta}} +
\theta_{\chi}^2\left|\frac{t-t_0}{T_{}}\right|^{-\delta} \right)^{-1}\cdot%
\left(1+\theta_{\chi}\left|\frac{t-t_0}{T_{}}\right|^{-\delta} \right)
\end{align}

The constants are subject to the following two sets of ``$SU(2)$-like'' constraints
\begin{align}
\bp_{\gamma} G\bp_{\gamma} &= 0 & \bp_{\gamma} \cdot \bd &= 1 & \delta &= -\bq_1\cdot\bp_i &
\bq_1& \cdot \bal_0 = \ln\left(\frac{q d^2 <\bq_1, \bq_1>}{8} \right) \\
\bp_{f}^{(\chi)} G\bp_f ^{(\chi)}&= 0 & \bp_{f}^{(\chi)}\cdot \bd &= 1 %
& \delta_{\beta} &= -\bq_2\cdot\bp_i & %
\bq_2&\cdot\left[\bal_0-\bp_i\ln \left|\frac{T}{T_{\beta}}\right|\right] = %
\ln\left(\frac{3d_{\chi}^2<\bq_2,\bq_2>}{4}\right) %
\end{align}
where $\gamma=i,f$, and the scalar product $<\cdot ,\cdot >$ is again defined by $<{\bf
a},{\bf b}> = {\bf a}^T G^{-1} {\bf b}$. Moreover, ${\bf p}_f$, ${\bf p}_f^{(\chi)}$and
${\bf p}_i$ are related by the two $SU(2)$ maps
\begin{align}
\bp_f - \bp_i = \delta \frac{2G^{-1}\bq_1}{<\bq_1 ,\bq_1 >} \quad, \quad %
\bp_f^{(\chi)} - \bp_i = \delta_{\beta} \frac{2G^{-1}\bq_2}{<\bq_2 ,\bq_2 >}
\end{align}

Finally, the fractional quantities $\theta_z,\theta_\chi$ are fixed according to
\begin{align}
\theta_z = \frac{\bq_1\cdot\bp_i}{(\bq_1+\bq_2)\cdot\bp_i} \quad, \quad %
\theta_\chi =  \frac{\bq_2\cdot\bp_i}{(\bq_1+\bq_2)\cdot\bp_i}
\end{align}

These satisfy $0\leq\theta_z,\theta_\chi \leq 1$ and $\theta_{z}+\theta_{\chi}=1$. For
clarity, we now present the solutions and constraints for $\bal$ in component fields. These
read

\begin{align}
\alpha-\alpha_0 =\frac{1}{3}\ln \left|\frac{t-t_0}{T}\right|&\\
%%%
\beta-\beta_0= p_{\beta,i}\ln\left|\frac{t-t_0}{T}\right| &+
(p_{\beta,f}-p_{\beta,i})\ln\left[ 1 + \left| \frac{t-t_0}{T} \right|^{-\delta}\left(1+
\theta_z^2 \left|\frac{t-t_{0}}{T_{\beta}}\right|^{-\delta_{\beta}}\right)\right]^{-1/\delta}\nonumber\\
&+ (p_{\beta,f}^{(\chi)}-p_{\beta,i})\ln \left[1 + %
\left|\frac{t-t_0}{T_{\beta}}\right|^{-\delta_{\beta}}%
\left(1+\theta_{\chi}^2 \left|\frac{t-t_{0}}{T}\right|^{-\delta}\right)\right]^{-1/\delta_{\beta}}\\
%%%
\phi-\phi_0= p_{\phi,i}\ln\left|\frac{t-t_0}{T}\right| &+ (p_{\phi,f}-p_{\phi,i})\ln\left[ 1
+ \left| \frac{t-t_0}{T} \right|^{-\delta}\left(1+ \theta_z^2
\left|\frac{t-t_{0}}{T_{\beta}}\right|^{-\delta_{\beta}}\right)\right]^{-1/\delta}\\
&+ (p_{\phi,f}^{(\chi)}-p_{\phi,i})\ln \left[1 + %
\left|\frac{t-t_0}{T_{\beta}}\right|^{-\delta_{\beta}}%
\left(1+\theta_{\chi}^2
\left|\frac{t-t_{0}}{T}\right|^{-\delta}\right)\right]^{-1/\delta_{\beta}}
\end{align}

The constants $\delta,p_{\beta,i},t_{0},T,d,z_0$ all occurred in the previous $SU(2)$
solutions and so are familiar. The three new constants are given by
$T_{\beta},\chi_0,d_{\chi}$ with the remainder constrained according to
\begin{align*}
\delta_{\beta}=-2p_{\beta,i} \quad, \quad
\theta_{z}=1-\theta_{\chi}=\frac{\delta}{\delta+\delta_{\beta}} \quad, \quad \beta_0=
\ln\left(2d_{\chi}\right)+p_{\beta,i}\ln\left|\frac{T}{T_{\beta}}\right| \quad,\quad
\beta_0-\phi_0= \ln\left(\frac{3}{2qd^2}\right)
\end{align*}
\begin{align*}
 \left(\begin{array}{c}
         p_{\beta ,f}\\
         p_{\phi ,f}
         \end{array}\right) =
 \frac{1}{2}\left( \begin{array}{rr}
                        1&1\\
                        3&-1
                    \end{array}\right)
 \left(\begin{array}{c}
       p_{\beta ,i}\\
       p_{\phi ,i}
       \end{array}\right) \quad, \quad
 \left(\begin{array}{c}
         p_{\beta ,f}^{(\chi)}\\
         p_{\phi ,f}^{(\chi)}
         \end{array}\right) =
          \left( \begin{array}{rr}
                        -1&0\\
                         0&1
                    \end{array}\right)
 \left(\begin{array}{c}
       p_{\beta ,i}\\
       p_{\phi ,i}
       \end{array}\right)
\end{align*}

Before discussing these $SU(3)$ solutions in more detail, we should also comment on the
solutions for the additional fields $\nu,\sigma$ satisfying Eq.~\eqref{eq:todatrunc}. It
transpires that the $\sigma$ solution involves a non-elementary integral, and so cannot be
presented analytically. However, its behaviour can always be computed numerically. On the
other hand, the field $\nu$ takes the simple form
\begin{align}\label{eq:su3nu}
\nu-\nu_0=d_{\nu} \left( 1+ \left|\frac{t-t_0}{T_{}}\right|^{\delta} +
\theta_{z}^2\left|\frac{t-t_0}{T_{\beta}}\right|^{-\delta_{\beta}} \right)^{-1}\cdot %
\left( \theta_{\nu}+\theta_{z}\left|\frac{t-t_0}{T_{\beta}}\right|^{-\delta_{\beta}} \right)
\end{align}
subject to the conditions
\begin{align}
d_{\nu}\equiv d\left(\chi_0+d_\chi\right) \quad, \quad \theta_\nu \equiv
\chi_0\left(\chi_0+d_\chi\right)^{-1}
\end{align}

Interestingly, the field $\nu$ can reverse field-velocity midway through its evolution, and
so ``turn-around'' or ``bounce''. This peculiar effect will have important ramifications
when we transform the $SU(3)$ solutions later on, for it will allow the brane field $z$ to
bounce as well.

\subsection{Analysis and Validity of the $SU(3)$ Toda Model}\label{subsec:su3ellipse}

There are two distinct $SU(2)$ models embedded non-trivially in these solutions. If we take
the limit $T_{\beta} \gg T$ then the early-time behaviour of the fields is formally
identical to the solutions Eq.~\eqref{eq:su2solns} of the previous section. If instead we
reverse the temporal sequence by choosing  $T_{\beta} \ll T$ then the early time behaviour
is another three-field $SU(2)$ model involving $\chi$ as the axion. These two models are
\emph{not} decoupled as they would be in the $SU(2)\times SU(2)$ Toda case, but instead are
non-trivially mixed inside the $SU(3)$ model. Only in extreme cases can we discern the
underlying $SU(2)$ components, and so at a general, intermediate time there will not be a
clean separation of the effects of the $z$ and $\chi$ motions. Indeed, these two embedded
behaviours couple and compete with one another, and attempt to drive the expansion powers
according to the two conflicting processes
\begin{align*} {\bf p}_i \rightarrow {\bf p}_f \quad, \quad {\bf
p}_{i} \rightarrow {\bf p}_{f}^{(\chi)}
\end{align*}
In general, this means that neither ${\bf p}_f$ nor ${\bf p}_{f}^{(\chi)}$ individually
succeed in becoming the actual expansion powers that the system adopts at late-time.
However, the system may spend some part of its evolution at \emph{intermediate}
rolling-radii states where it adopts these powers temporarily. The true late-time
rolling-radii states ${\bf p}_{f}'$ are in fact determined from a ``combined'' relation
given by
\begin{align}~\label{su3endstates}
\bp_{f}' - \bp_i = (\delta+\delta_{\beta}) \frac{2G^{-1}\bq}{<\bq ,\bq >} \quad, \quad  \bq=
\bq_1+\bq_2
\end{align}
Notice that the final states are computed \emph{as if} the system followed an ordinary
$SU(2)$ model with a combined parameter $\delta+\delta_{\beta}$. However, the intermediate
behaviour strongly deviates from any such simple $SU(2)$ evolution, and we should treat
Eq.~\eqref{su3endstates} merely as a formal tool for deducing the rolling-radii endpoints of
the trajectory.

\begin{figure}[htbp]
\begin{center}
\includegraphics[angle=-90,width=8cm]{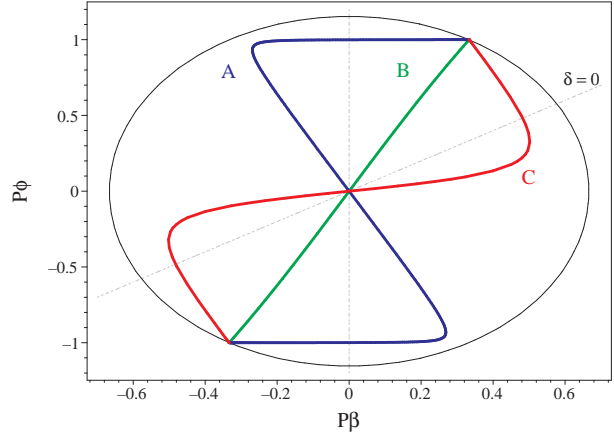}
\caption[$SU(3)$ case 3 behaviour]{\label{fig:SU3ellipse}\textit{ A typical set of $SU(3)$
trajectories. Curve  \blue{A} has two horizontal lines and one diagonal line, and so
contains two $\chi$ displacements and one $z$ displacement. Curve \green{B} represents a
special, degenerate case for which both $z$ and $\chi$ evolve at once and mimic a single
field. Curve \red{C} has two diagonal lines and one horizontal line, and so contains two $z$
motions separated by a $\chi$ displacement. All intermediate cases between these curves are
possible. }}
\end{center}
\end{figure}

\begin{figure}[htbp]
\begin{center}
\includegraphics[angle=-90,width=7cm]{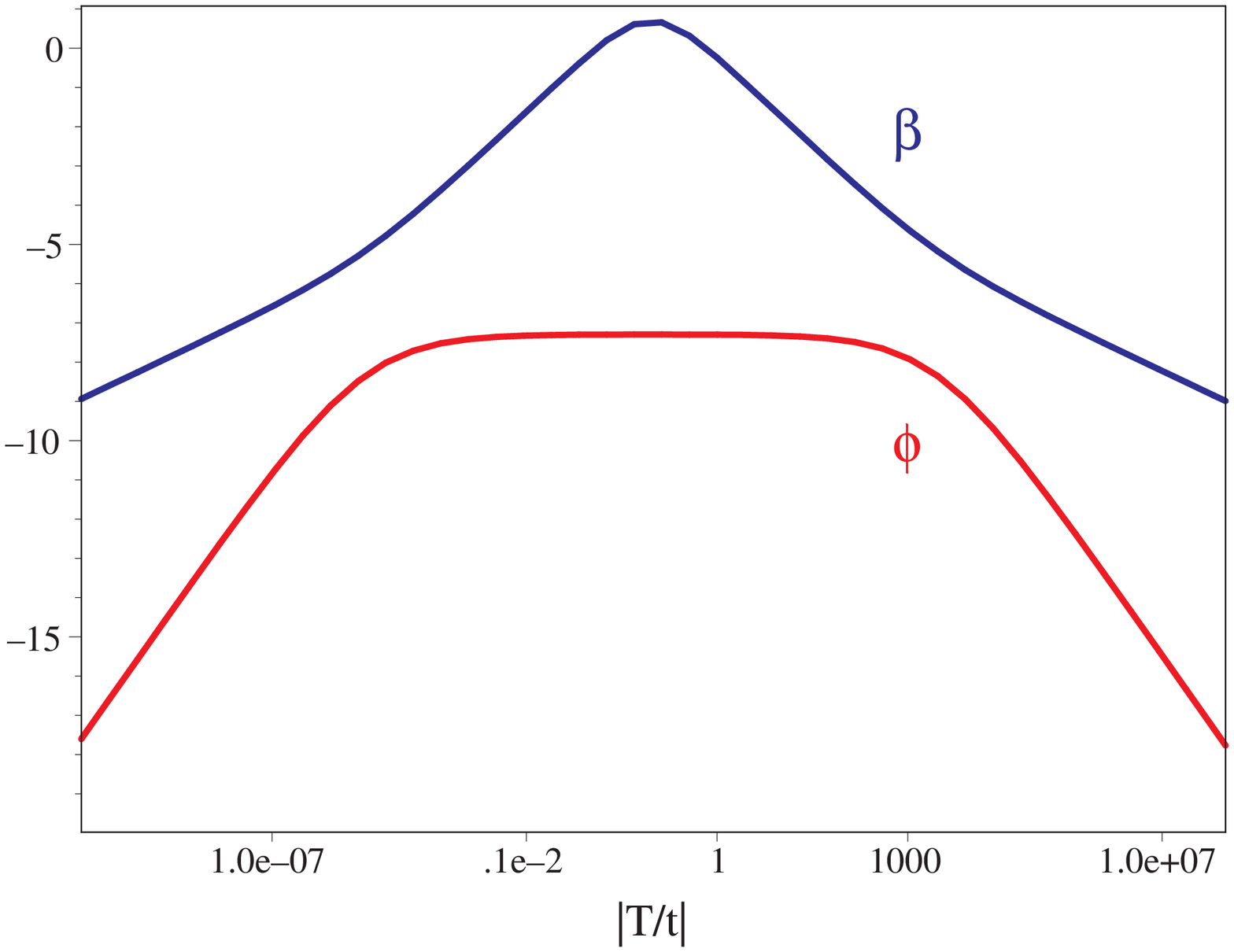}
\includegraphics[angle=-90,width=7cm]{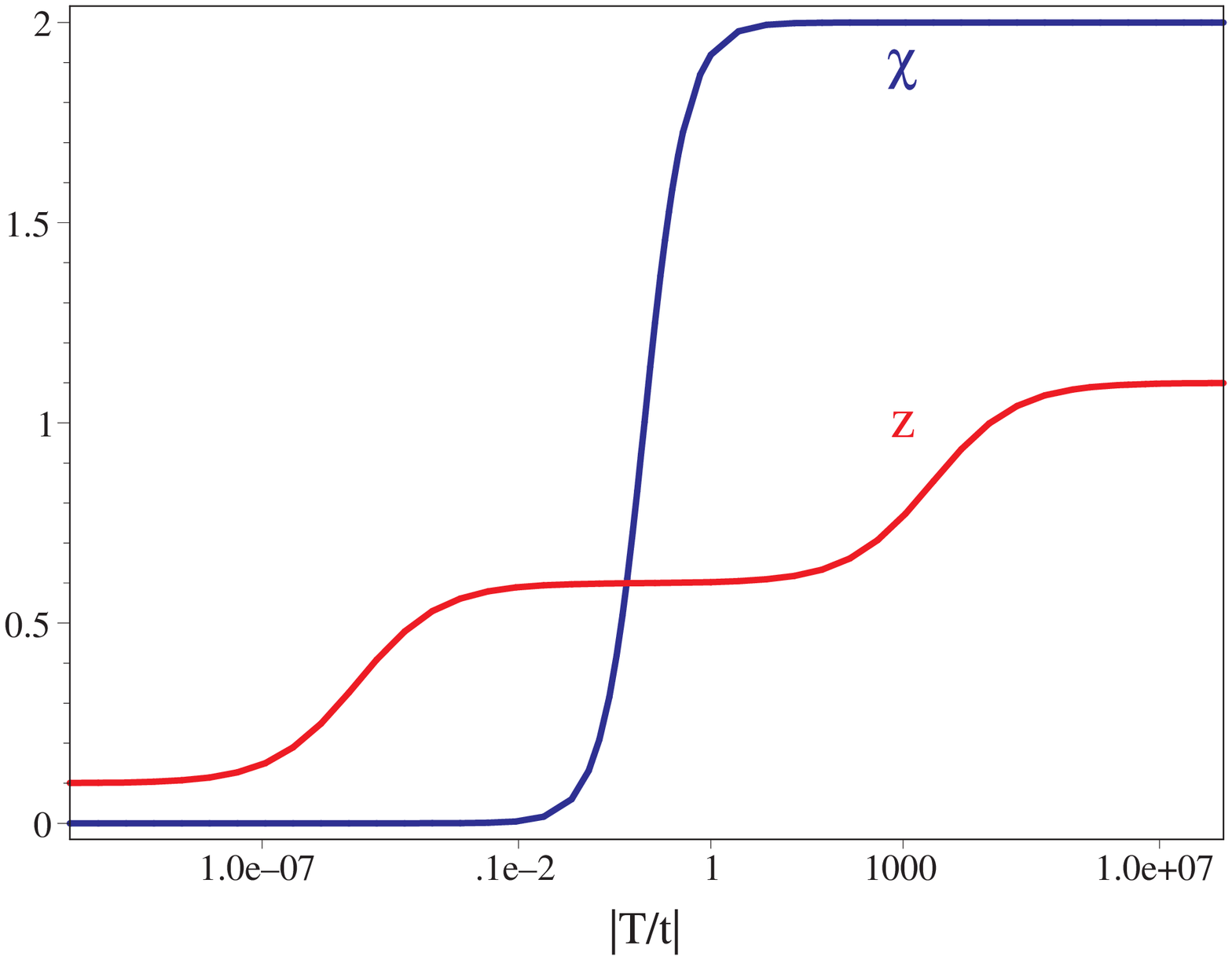}
\caption[$SU(3)$ Toda Case 1 field behaviour]{\label{fig:SU3fields} \textit{ An interesting
example of the field evolution. The left hand plot shows the  fields \blue{$\beta$} and
\red{$\phi$}. The kinks in these curves are caused by the displacements of \red{$z$} and
\blue{$\chi$}, which are shown in the right hand plot. Notice that $z$ moves twice in
succession. }}
\end{center}
\end{figure}

To see this clearly, we plot the field behaviour on the ellipse as we did in
Section~\ref{sec:su2} (see Fig.~\ref{fig:SU3ellipse}). Again, the solutions break in $(\pm)$
branches, with $\mathbf{p}_i$ corresponding to the early-time expansion powers only if they
simultaneously satisfy the two inequalities
\begin{align}
\delta,\delta_{\beta} > 0  \quad {\rm on} \quad (-) \quad, \quad %
\delta,\delta_{\beta}< 0  \quad {\rm on} \quad (+)
\end{align}
The additional $\delta_{\beta}$ condition restricts the accessible early-time powers to a
narrower range of states compared to the $SU(2)$ model. In Fig.~\ref{fig:SU3fields} we also
plot the displacements of $z$ and the $\chi$ axion. This illustrates the important fact that
the fields $z,\chi$ can undergo \emph{two} successive displacements, since each is coupled
to the time-development of the other.

\newpage
To complete this section, we now comment on the validity of these $SU(3)$ Toda solutions.
One can show that
\begin{align}~\label{eq:su3couple}
e^{\beta-\phi}=\frac{3}{2qd^2}\left|\frac{t-t_0}{T}\right|^{\delta}\cdot %
\frac{\left[ 1 + \left| \frac{t-t_0}{T_{}} \right|^{-\delta}\left(1+ \theta_z^2
\left|\frac{t-t_{0}}{T_{\beta}}\right|^{-\delta_{\beta}}\right)\right]^2}{%
1 +\left|\frac{t-t_0}{T_{\beta}}\right|^{-\delta_{\beta}}%
\left(1+\theta_{\chi}^2 \left|\frac{t-t_{0}}{T_{}}\right|^{-\delta}\right)}
\end{align}
Using this, one can easily verify that in the asymptotic limits
\begin{align*}
\epsilon_{k} \sim \left|\frac{t-t_0}{T}\right|^{+\delta} \rightarrow \infty \quad \text{ at
early time} \quad, \quad %
\epsilon_{k} \sim \left|\frac{t-t_0}{T_{\beta}}\right|^{-\delta_{\beta}} \rightarrow \infty
\quad \text{ at late time}
\end{align*}
Notice that this follows only because $\delta,\delta_{\beta}$ are both positive (negative)
on the negative (positive) time-branch. Consequently, the $SU(3)$ solutions cannot be
trusted asymptotically, as with the previous $SU(2)$ solutions of Section~\ref{sec:su2}.
Further investigation of Eq.~\eqref{eq:su3couple} also reveals that it can never be made
smaller than the leading coefficient, which is of order $1/qd^2$. This demonstrates that the
smallest attainable value of the $\epsilon_{k}$ is given by
\begin{align}
\epsilon_{k}|_{\mathrm{min}} \sim \frac{1}{d^2}
\end{align}
Hence, to achieve $\epsilon_{k} \ll 1$ the solutions require us to take $d \gg 1$ and allow
the brane to leave the orbifold interval. As such, the $SU(3)$ model has similar problems to
the $SU(2)$ model. Of course, as long as we are interested in relatively short timescales,
and always concentrate on the brane behaviour inside of the interval but away from the
boundaries (and the collision) then no particular problem is posed. Away from the boundaries
the $SU(3)$ solutions with $d \gg 1$ are reliable for a short time, and the fact that the
brane must \emph{eventually} leave the interval does not change this fact. Hence, there are
always regions where all fields are evolving in an $\epsilon_{k} \ll 1$ regime with the
brane inside the interval.

However, it would obviously be valuable if these regions could be extended to cover the
entire displacement profile of the brane, such that the brane moves and comes to rest while
remaining inside the interval with $\epsilon_{k} \ll 1$ throughout. Although this is
impossible with the $SU(3)$ solutions themselves, when we come to symmetry-transform the
$SU(3)$ solutions we will find circumstances under which $d<1$ and $\epsilon_{k} \ll 1 $
simultaneously.
%=================================SYMMETRIES=============================================================

\section{Application of the Symmetries}\label{sec:symmetries}

We now apply the symmetries presented in our companion paper Ref.~\cite{Copeland:2005mk} to
the $SU(2)$ and $SU(3)$ models in turn. (For previous work in this area, also see
Ref.~\cite{deCarlos:2004yv}). These symmetries mix the scalar fields together in new
combinations, and yet leave the action Eq.~\eqref{4daction} invariant. Consequently, the new
time-dependent combinations for the fields that emerge, no matter how complicated, still
solve the equations of motion. The seven-dimensional symmetry group $G$ is a maximal
parabolic subgroup of $Sp(4,\mathbb{R})$ given by
\begin{align}~\label{eq:symmgroup}
G = SL(2,\mathbb{R}) \ltimes SU(1,2)/U(2)
\end{align}
where $\ltimes$ denotes a semidirect product. The scalar-field space is then the group
manifold $\mathcal{M}\cong Sp(4,\mathbb{R})/U(2)$, but equipped with a nonhomogeneous
Riemannian metric such that the total symmetry group fills out only $G$ and not the whole of
$Sp(4,\mathbb{R})$. There are seven, distinct types of transformations $L^{i}$ that can be
applied to the scalar fields, each one controlled by a continuously adjustable real constant
$c_{i}$. These are given by

\begin{align}~\label{finitetrans}
L^{1}&: \quad \beta \rightarrow \beta + c_{1} && \chi \rightarrow \chi \:e^{c_{1}}
&&  z \rightarrow z \: e^{-c_{1}/2} && \nu  \rightarrow \nu \:e^{c_{1}/2}& \nonumber\\
L^{2}&: \quad e^{\beta} \rightarrow
\frac{e^{\beta}}{(1+c_{2}\chi)^{2}+\frac{1}{4}c_{2}^{2}e^{2\beta}}
 && \chi \rightarrow \frac{\chi(1+c_{2}\chi) + \frac{1}{4}c_{2}e^{2\beta}}{ %
\quad (1+c_{2}\chi)^{2}+\frac{1}{4}c_{2}^{2}e^{2\beta}}
 && \sigma \rightarrow \sigma- c_{2} \cdot 2q\nu^2 && z \rightarrow z + c_{2}\nu & \nonumber \\
L^{3}&: \quad \chi \rightarrow \chi + c_{3} &&  \sigma \rightarrow \sigma - c_{3} \cdot
2qz^{2} && \nu \rightarrow \nu + c_{3}z  &&   & \nonumber\\
L^{4}&: \quad \phi  \rightarrow \phi + c_{4} && \sigma \rightarrow \sigma \:e^{c_{4}} && z
\rightarrow z \: e^{c_{4}/2} && \nu \rightarrow \nu \:e^{c_{4}/2} & \nonumber \\
L^{5}&: \quad \sigma \rightarrow \sigma -4q\nu\cdot c_{5} &&  z \rightarrow z + c_{5} && &&
& \nonumber\\
L^{6}&: \quad \sigma \rightarrow \sigma +4qc_{6} &&   &&   &&   & \nonumber\\
L^{7}&: \quad \nu \rightarrow \nu + c_{7} &&   &&  && &
\end{align}

Some of these transformations will simply amount to reparameterisations of the existing
integration constants in the solutions they are applied to. Some, however, change the
solutions into new functional forms, which will in turn be completely new solutions to the
equations of motion. In particular, note that $L^{1},L^{2},L^{3}$ represent the
$SL(2,\mathbb{R})$ group of transformations, which act on $S,T,Z$ as follows
\begin{align}\label{eq:Tduality}
T'= \frac{aT-ib}{icT+d} \quad,\quad Z'= \frac{Z}{icT+d} \quad, \quad S'=S-
\frac{icqZ^2}{icT+1}\quad \mathrm{with} \quad  a,b,c,d \in \mathbb{R} \quad, \quad ad-bc=1
\end{align}
Here the four constants $a,b,c,d$ (subject to one constraint) are proportional to
$c_{1},c_{2},c_{3}$, and are better adapted to the $SL(2,\mathbb{R})$ symmetry. In
particular, they can be considered the four entries of a $2\times 2$ $SL(2,\mathbb{R})$
matrix. Having now rewritten the $SL(2,\mathbb{R})$ transformations in this compact form,
note the crucial fact that Eq.~\eqref{eq:Tduality} does \emph{not} represent ordinary
$T$-duality, for the complex coordinates $S$ and $Z$ must also be transformed. Nonetheless,
the action on $T$ alone is indistinguishable from conventional $T$-duality, and so all-told
we will dub this a ``generalised'' $T$-duality. These generalised $T$-duality
transformations can produce exceedingly complicated new behaviours, and can significantly
affect any existing time-dependent solutions that they are applied to. Consequently, we can
expect to derive new solutions to the equations of motion by transforming the $SU(2)$ and
$SU(3)$ models using these symmetries. While it was shown in Ref.~\cite{Copeland:2005mk}
that the symmetries do not form a transitive group on $\mathcal{M}$, so that we cannot use
them to build the \emph{general} solution to the equations of motion, we can nonetheless
make significant progress in this direction.

%============================SU2 TRANSFORM ===================================================================
\section{Transforming the $SU(2)$ model}\label{subsec:su2trans}

We now apply the finite symmetries $L^{i}$ to the $SU(2)$ model. Only $L^{2},L^{3}$ have a
non-trivial effect, and they modify the system such that it is no longer a simple Toda model
that can be solved using the Toda methodology. This, of course, is the crucial reason why we
use the symmetries in the first place, as they allow us access to complicated new solutions
that we cannot otherwise uncover using standard methods. Although the brane only undergoes
one displacement, we will find that the $\epsilon_{k}$ parameters can have significantly
different development in these transformed solutions. Specifically, in certain cases the
$\epsilon_{k}$ are naturally decreasing into the past or future.

\subsection{Transformed-$SU(2)$ Solutions}

One can verify that the symmetries transform the $SU(2)$ solutions into the following form
\begin{align}~\label{eq:su2transsolns}
\bal-\bal_0 = {\bf p}_i \ln\left| \frac{t-t_0}{T} \right| & + \left( \tilde{{\bf p}}_f
-{\bf p}_i \right) \ln \left( 1 + \left| \frac{t-t_0}{T} \right|^{-\tilde{\delta}}\right)^{-1/\tilde{\delta}}\nonumber\\
+&\left( {\bf p}_{f}^{(\chi)} -{\bf p}_{i} \right) %
\ln \left\{\left|\frac{t-t_0}{T_{\beta}} \right|^{-\Delta \delta_{\beta}}\left[1 +
\left|\frac{t-t_0}{T_{\beta}}
\right|^{-s\delta_{\beta}}\left(1+\left|\frac{t-t_{0}}{T}\right|^{-\tilde{\delta}}\right)\right]\right\}^{-1/\delta_{\beta}}\\
z-z_0&=d \left( 1+ \left|\frac{t-t_0}{T_{}}\right|^{\tilde{\delta}}\right)^{-1}\\
\chi-\chi_0 &= d_{\chi} \left[ 1+
\left|\frac{t-t_0}{T_{\beta}}\right|^{s\delta_{\beta}}\left(1 +
\left|\frac{t-t_0}{T_{}}\right|^{-\tilde{\delta}} \right)^{-1} \right]^{-1}\\
\sigma-\sigma_0 &= - 2q\chi_{0}\left[z_{0} + d\left( 1 + \left|\tT\right|^{\tilde{\delta}}
\right)^{-1} \right]^{2}\\
\nu-\nu_{0} &=0
\end{align}
Here we have defined the combinations
\begin{align*}
s=\pm 1 \quad, \quad 2\Delta= 1-s \quad, \quad %
\tilde{\delta} = \delta + \Delta \delta_{\beta}
\end{align*}
We are, however, not free to pick $s$ in an arbitrary fashion, as the choice of sign
crucially depends on the choice of initial expansion powers. The permissible choices are
listed in table~\ref{table:su2trans}.
\begin{table}[htbp]
\begin{center}
\begin{tabular}{|c|c|c|}
\hline Expansion power  & $\delta,\delta_{\beta}<0$ on $(+)$  & $\delta+\delta_{\beta}<0,\delta_{\beta}<0,\delta>0$ on $(+)$\\[2pt]
range & $\delta,\delta_{\beta}>0$ on $(-)$ &
$\delta+\delta_{\beta}>0,\delta_{\beta}>0,\delta<0$
on $(-)$ \\[6pt]
\hline Allowed choices  & $s=\pm 1$ & $s=-1$ \\[5pt] \hline
\end{tabular}
\caption[]{\label{table:su2trans} Table to show which of the values $s=\pm 1$ are valid
choices given a set of initial expansion powers. Note that $s=+1$ is not a valid choice in
the second row of the table. }
\end{center}
\end{table}

The remaining constants are then subject to the same constraints as in the $SU(3)$ model,
namely
\begin{align}
\bp_{\gamma} G\bp_{\gamma} &= 0 & \bp_{\gamma} \cdot \bd &= 1 & \delta &= -\bq_1\cdot\bp_i &
\bq_1& \cdot \bal_0 = \ln\left(\frac{q d^2 <\bq_1, \bq_1>}{8} \right) \\
\bp_{f}^{(\chi)} G\bp_f ^{(\chi)}&= 0 & \bp_{f}^{(\chi)}\cdot \bd &= 1 %
& \delta_{\beta} &= -\bq_2\cdot\bp_i & %
\bq_2&\cdot\left[\bal_0-\bp_i\ln \left|\frac{T}{T_{\beta}}\right|\right] = %
\ln\left(\frac{3d_{\chi}^2<\bq_2,\bq_2>}{4}\right) %
\end{align}
where $\gamma=i,f$ and $\tilde{{\bf p}}_f$, ${\bf p}_f^{(\chi)}$and ${\bf p}_i$ are related
by the two $SU(2)$ maps
\begin{align}
\bp_f^{(\chi)} - \bp_i = \delta_{\beta}
\frac{2G^{-1}\bq_2}{<\bq_2 ,\bq_2 >}\quad, \quad %
\tilde{\bp}_f - \bp_i = \tilde{\delta} \frac{2G^{-1}\tilde{\bq}}{<\tilde{\bq} ,\tilde{\bq} >} %
\quad, \quad \tilde{\bq} = \bq_1 + \Delta \bq_2
\end{align}

Notice the crucial fact that the system does \emph{not} necessarily have the same asymptotic
behaviour as the $SU(2)$ model. To see this, we note that the early-time powers $\bp_{i}$
can now be taken from anywhere in the region
\begin{align*}
\delta+\delta_{\beta}, \delta_{\beta}<0 \quad \mathrm{on} \quad (+) \quad,\quad %
\delta+\delta_{\beta}, \delta_{\beta}>0 \quad \mathrm{on}\quad (-)
\end{align*}
Most of this region was unavailable in the original $SU(2)$ model, and so the permissible
asymptotic behaviours have expanded into completely new regions. This is very different to
the $SU(3)$ model, which consistently \emph{narrowed} the range of powers compared to the
old $SU(2)$ case, but did \emph{not} expand the allowed range of powers at all. These newly
accessible regions are entirely a consequence of the symmetry transformations, whose effects
were entirely absent in the original $SU(2)$ and $SU(3)$ cases. Therefore, we can anticipate
entirely different behaviour for the $\epsilon_k$ parameters in the asymptotic limits.

For clarity, we now present the component field representation of $\bal$:
\begin{align}
\alpha-\alpha_0=\frac{1}{3}\ln \left|\frac{t-t_0}{T}\right|&\\
\beta-\beta_0= p_{\beta ,i} \ln\left|\tT\right| %
&+(p_{\beta,f}-p_{\beta,i})\ln\left(1+ \left|\tT\right|^{-\tilde{\delta}}\right)^{-1/\tilde{\delta}}\\
&+(p_{\beta,f}^{(\chi)}-p_{\beta,i})\ln \left\{\left|\frac{t-t_0}{T_{\beta}}
\right|^{\delta-\tilde{\delta}}\left[1 + \left|\frac{t-t_0}{T_{\beta}}
\right|^{-\delta_{\beta}}\left(1+\left|\frac{t-t_{0}}{T}\right|^{-\tilde{\delta}}\right)\right]\right\}^{-1/\delta_{\beta}} \nonumber \\
\phi-\phi_0= p_{\phi ,i} \ln\left|\tT\right| %
&+(p_{\phi,f}-p_{\phi,i})\ln\left(1+
\left|\tT\right|^{-\tilde{\delta}}\right)^{-1/\tilde{\delta}}
\end{align}
subject to the constraints
\begin{align*}
\delta=p_{\beta,i}-p_{\phi,i} \quad, \quad \delta_{\beta}=-2p_{\beta,i} \quad, \quad
\beta_0= \ln\left(2d_{\chi}\right)+p_{\beta,i}\ln\left|\frac{T}{T_{\beta}}\right|
\quad,\quad \beta_0-\phi_0= \ln\left(\frac{3}{2qd^2}\right)
\end{align*}
\begin{align*}
 \left(\begin{array}{c}
         \tilde{p}_{\beta ,f}\\
         \tilde{p}_{\phi ,f}
         \end{array}\right) =
          \frac{1}{2}\left( \begin{array}{cc}
                          1 & s\\
                         3s & -1
                    \end{array}\right)
 \left(\begin{array}{c}
       p_{\beta ,i}\\
       p_{\phi ,i}
       \end{array}\right)
\quad, \quad
\left(\begin{array}{c}
         p_{\beta ,f}^{(\chi)}\\
         p_{\phi ,f}^{(\chi)}
         \end{array}\right) =
          \left( \begin{array}{rr}
                        -1&0\\
                         0&1
                    \end{array}\right)
 \left(\begin{array}{c}
       p_{\beta ,i}\\
       p_{\phi ,i}
       \end{array}\right)
\end{align*}

As in all the cases considered, the expansion powers $(p_{\beta,i},p_{\phi,i})$ are
constrained to the ellipse as defined in Eq.~\eqref{eq:ellipse}. In combination with the
constraints above, this automatically forces $\beta$ and $\phi$ to be in rolling-radii
regimes at late-time that are also on the ellipse.

\subsection{Analysis and validity of the transformed-$SU(2)$ model}\label{subsec:su2transellipse}

\begin{figure}[htbp]
\begin{center}
\includegraphics[angle=-90,width=8cm]{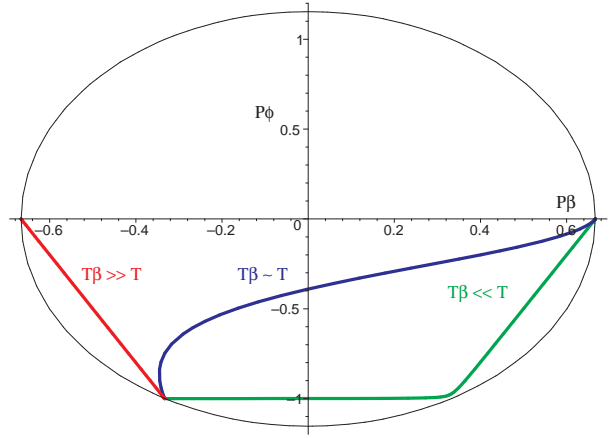}
\caption[$SU(2)$ transformed class 1]{\label{fig:SU2transcase1} \textit{ Transformed $SU(2)$
ellipse behaviour for  \red{$T_{\beta} \gg T$}, \blue{$T_{\beta} \sim T$}, \green{$T_{\beta}
\ll T$}, plotted on the $(-)$ branch. The trajectories begin at the lower left as $t-t_{0}
\rightarrow -\infty$, and evolve to the upper right as $t-t_{0} \rightarrow 0$. }}
\end{center}
\end{figure}

As in the previous section, we plot a particular example of the $\beta,\phi$ evolution
across the ellipse on the $(-)$ branch (see Fig~\ref{fig:SU2transcase1}). The behaviour
breaks down into three generic cases based on the relative magnitudes of the timescales $T$
and $T_{\beta}$. Notice that, irrespective of these magnitudes, the field $z$ can only ever
undergo one displacement, and so behaves in a manner identical to the old $SU(2)$ case.
However, the crucial thing is that we can now achieve the same $SU(2)$ behaviour for $z$
inside a set of solutions that have completely different development for the $\epsilon_k$.
In the particular example given, the originally diverging values of $\epsilon_k$ at
late-time are now decreasing to arbitrarily small values into the future. Thus, the
solutions become more and more reliable into the future. This is in stark contrast to the
$SU(2)$ model from which they originated, and demonstrates that the new $\chi$ behaviour is
crucial in suppressing gravitational corrections to the four-dimensional theory.

%==================================================================================================== FIGURE

\begin{figure}[htbp]
\begin{center}
\includegraphics[angle=-90,width=7cm]{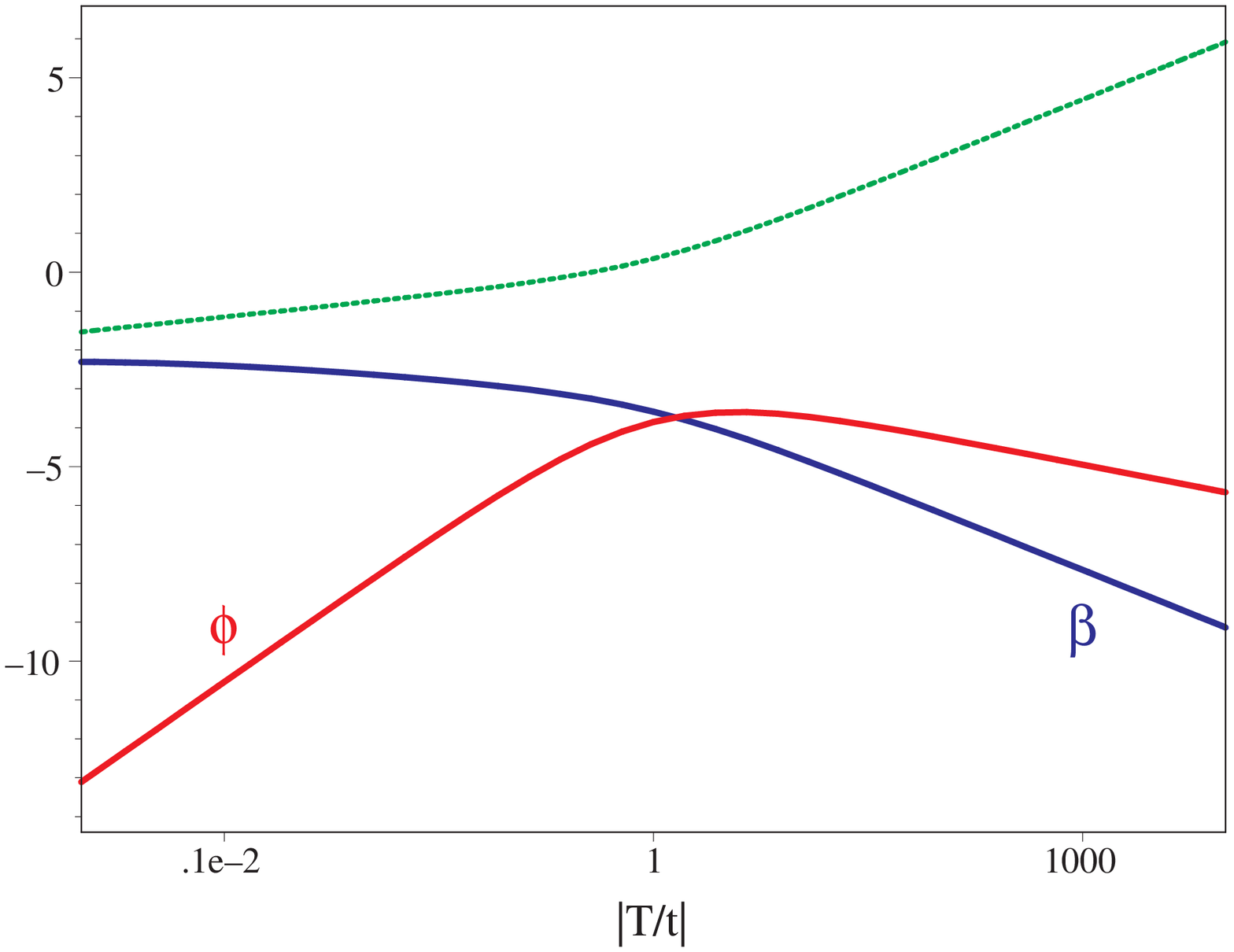}
\includegraphics[angle=-90,width=7cm]{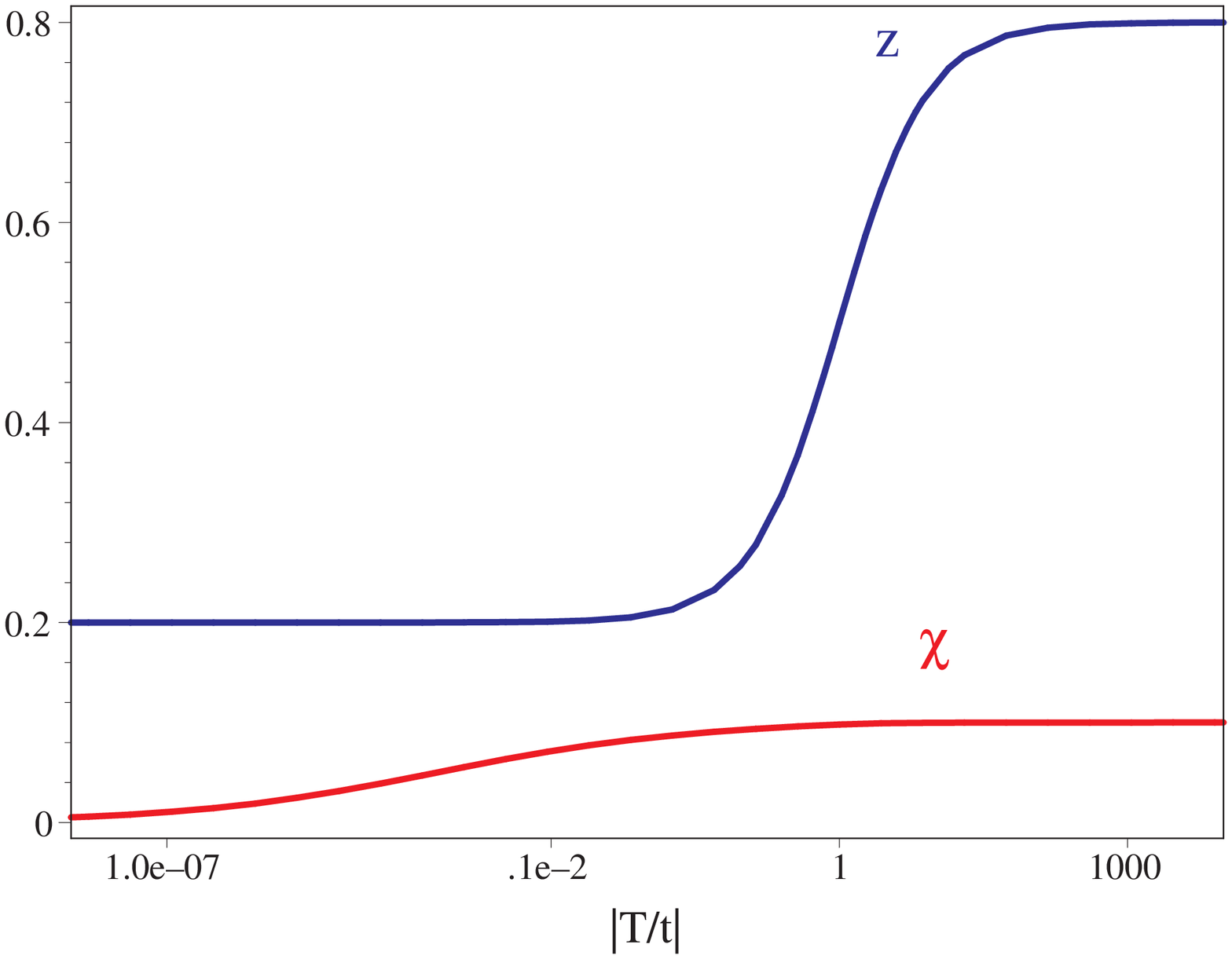}
\caption[$SU(2)$ transform class 1 field behaviour]{ \textit{In the left hand plot we see
\red{$\phi$} and the $SU(2)$-transformed solution for \blue{$\beta$}, with the remaining
curve corresponding to the original untransformed $SU(2)$ solution for $\beta$. In the right
hand plot we see the distinct \blue{$z$} and \red{$\chi$} displacements, with the motion of
the bulk brane $z$ occurring second. This allows the brane to displace with $d<1$ and yet
still be in a $\epsilon_{k} \ll 1 $ regime. \label{fig:SU2transABfields} }}
\end{center}
\end{figure}

Moreover, in Fig.~\ref{fig:SU2transABfields} we plot the $T_{\beta} \ll T$ field behaviour
of $z,\chi$, such that the $z$ displacement occurs in the reliable $\epsilon_{k} \ll 1 $
regime \emph{after} the change in $\chi$. Crucially, this means that the brane motion can
occur in a $\e_{k} \ll 1$ region \emph{without} requiring $d>1$. Such behaviour could never
have occurred in the original $SU(2)$ model, and is a consequence of the manner in which the
fields $z$ and $\chi$ are incorporated together into a new, global structure for the overall
solutions.

%===================================================================================================FIGURE

The fact that the displacement of $z$ can be made to entirely occur in a $\epsilon_{k} \ll 1
$ regime, without requiring $d \gg 1$, constitutes a significant improvement over the
original $SU(2)$ model. This proves that a reliable solution for $z$ does not necessarily
require it to eventually leave the compact space . However, there are obviously other
possible examples beyond those shown in
Figs.~\ref{fig:SU2transcase1}-\ref{fig:SU2transABfields}, and we should now clarify the
precise circumstances in which the $\epsilon_{k}$ can be made to decrease. Once again we
consider the functional form of $\exp(\beta-\phi)$:
\begin{align}~\label{eq:su2transcouple}
e^{\beta-\phi}=\frac{3}{2qd^2}\left|\frac{t-t_0}{T}\right|^{\delta}\cdot %
\frac{\left( 1 + \left| \frac{t-t_0}{T_{}}
\right|^{-\tilde{\delta}}\right)^2}{%
\left|\frac{t-t_0}{T_{\beta}} \right|^{\delta-\tilde{\delta}}\left[1 +
\left|\frac{t-t_0}{T_{\beta}}
\right|^{\delta_{\beta}}\left(1+\left|\frac{t-t_{0}}{T}\right|^{-\tilde{\delta}}\right)\right]}
\end{align}
By choosing signs appropriately, either the early \emph{or} the late time limit can
become ``weakly-coupled'' with $\epsilon_{k} \ll 1$. However, only \emph{one} of the
asymptotic limits can be weakly-coupled, with the other still becoming ``strongly-coupled''
with $\epsilon_{k} \gg 1$. There are, of course, still solutions where $\epsilon_{k} \gg 1$
is attained in both limits. The full state of affairs is summarised in
table~\ref{table:su3transcouple}.

\begin{table}[htbp]
\begin{center}
\begin{tabular}{|c|c|c|}
\hline %
$s=+1$ & $s=-1$ & $s=-1$ \\[10pt]
\hline
$\delta,\delta_{\beta}<0$ on $(+)$
&$\delta,\delta_{\beta}<0$ on $(+)$  & $\delta+\delta_{\beta}<0,\delta_{\beta}<0,\delta>0$ on $(+)$\\[6pt]
$\delta,\delta_{\beta}>0$ on $(-)$
&  $\delta,\delta_{\beta}>0$ on $(-)$ & $\delta+\delta_{\beta}>0,\delta_{\beta}>0,\delta<0$ on $(-)$ \\[6pt]
\hline SW & SS & WS \\[6pt]
\hline
\end{tabular}
\caption[]{ \label{table:su3transcouple}\textit{Table to show the asymptotic values of the
coupling parameters $\epsilon_{k}$, depending on the sign of $s$. The notation is as
follows: strong-strong (SS), strong-weak (SW) and weak-strong (WS), where the first word
corresponds to the $\epsilon_{k}$ values in the early-time limit, and the second refers to
their values in the late-time limit. }}
\end{center}
\end{table}

Thus, there are three distinct types of solutions: weak-strong (WS), strong-weak (SW) and
strong-strong (SS). In all three cases we can arrange for the $z$ motion to occur in a
$\e_{k} \ll 1$ region with $d<1$. To do this, we simply recognise that at $t-t_{0} \approx
T$ we can always take $T_{\beta}\ll T$, and this will decrease the values of the
$\epsilon_{k}$ below 1 without requiring $d \gg 1$. Consequently, there is a tremendous
degree of flexibility in the solutions, and cases with $\epsilon_{k} \ll 1$ and $d<1$ are
quite generic.

Before leaving this section, we should also comment on stringy $\alpha'$ corrections. These
become strong as we probe small length scales at $\beta \sim 0$, and so encounter new
physics not accounted for in the effective supergravity description. As such, one must
always ensure that $\beta \gg 0$ to trust any supergravity solution. We note that this is
always possible for certain periods of time by an appropriate choice of integration
constants, and so there is no obstruction to finding regimes where $\epsilon_{k} \ll 1$ and
$\alpha'$ corrections are extremely small. The transformed $SU(2)$ solutions thus
incorporate all of the $z$ behaviour from the $SU(2)$ model, but now allow it to be
compressed inside of the orbifold interval whilst simultaneously suppressing all unwanted
corrections.
%============================================SU3 TRANSFORM =======================================================

\section{Transforming the $SU(3)$ Toda model}\label{subsec:su3trans}

We now apply the symmetries $L^{i}$ to the $SU(3)$ solutions, and so find a further class of
new solutions. One finds in this case that only the action of the $L^{2}$ transformation can
ever lead to new behaviour. This can be easily understood by noting that all the other
transformations leave the $SU(3)$ truncation conditions Eq.~\eqref{eq:todatrunc} invariant,
while $L^{2}$ allows $\dot{\nu}-\chi\dot{z}$ to become non-zero. The ``activation'' of this
combination takes us outside of the original $SU(3)$ Toda model, and into a new situation
that is not itself solvable by Toda methods. Nonetheless, the symmetries allow us to access
an exact, analytical description of the behaviour when this combination is non-zero. We will
find that the brane can undergo two displacements in opposite directions, and so reverse
direction without the presence of any explicit potentials. We will often call this a
``bouncing'' solution.

\subsection{Transformed-$SU(3)$ Solutions}

These new solutions, although exact, are complicated and difficult to present in an elegant
fashion. One means of presentation is to utilise two time-dependent functions $p,r$ that are
implicitly defined via the relations
\begin{align}
4r(4+p^2)^{-1}=e^{\beta}|_{SU(3)} \quad, \quad pr(4+p^2)^{-1} = \chi|_{SU(3)}
\end{align}
These are built out of the $\beta,\chi$ solutions from the old (untransformed) $SU(3)$
model. The \emph{new} transformed-$SU(3)$ solutions can then be written in the form
\begin{align*}
\alpha =\alpha|_{SU(3)} \quad,\quad \phi = \phi|_{SU(3)} \quad,\quad \nu = \nu|_{SU(3)}
\end{align*}
\begin{align}\label{eq:su3trans}
\beta = \ln \left\{4r \left[ 4+\left(p+c_{2}r\right)^2\right]^{-1}\right\} \quad,\quad %
\chi = r\left(p+c_{2}r\right)\left[4+\left(p+c_{2}r\right)^2\right]^{-1} \quad, \quad  %
z = z|_{SU(3)} + c_{2} \nu|_{SU(3)} &
\end{align}
Here $c_{2}$ is the real constant associated to the $L^{2}$ symmetry, and so corresponds to
a new integration constant that can be varied at will.  The remaining constants, it should
be emphasised, are taken from the original $SU(3)$ model, and we should treat their values
as determining an embedding of the old $SU(3)$ behaviour inside the newly transformed
solutions. Indeed, the fields $\alpha,\phi,\nu$ are unaffected by the transformations, and
evolve as in the old $SU(3)$ case in any event.

Notice as well that we cannot present $\sigma$ analytically, due to the fact that the
corresponding $SU(3)$ solution can only be computed numerically. As such, the transformed
$\sigma$ solution must also be computed numerically. However, we emphasise that these
numerical computations can be readily carried out with no obstruction, and that the symmetry
transformations induce perfectly sensible behaviour for $\sigma$ in all cases. In addition,
$\sigma$ can have no bearing on the time-development of the other fields, as the condition
$\dot{\sigma}+4qz\dot{\nu} = 0$ is preserved under the symmetry group
Eq.~\eqref{eq:symmgroup}. This means that $\sigma$ never appears in the equations of motion
of the other fields, and so can never induce any changes to the brane or remaining axions.
Consequently, we will not particularly concern ourselves with $\sigma$ from this point on.

Having derived these new solutions by applying the symmetries, we must now consider the
ramifications for the various fields involved. In particular, we are most interested in the
field $z$ and the issue of whether we can now achieve a sensible displacement at
weak-coupling. In the original $SU(3)$ model the field $z$ could undergo two successive
displacements in the same direction, but it could not do so whilst entirely within a
$\epsilon_{k} \ll 1$ regime. In the above case, however, the new field $z$ is an additive
mixture of the old $SU(3)$ behaviours for $z,\nu$. This creates a significant new level of
flexibility, and in the next section we will investigate the consequences for the brane and
its displacements.

\subsection{Analysis and Validity of the Transformed-$SU(3)$ model}\label{subsec:su3transellipse}

\begin{figure}[htbp]
\begin{center}
\includegraphics[angle=-90,width=8cm]{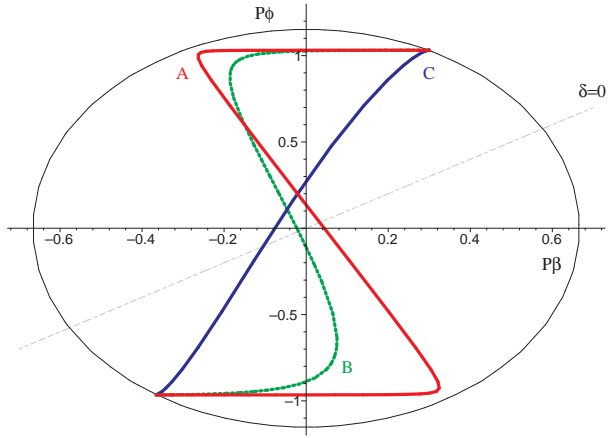}
\caption[$SU(3)$ transformed case 2 behaviour]{\label{fig:SU3trans2}\textit{ Here we show
some examples of the transformed-$SU(3)$ solutions. Curve \red{A} is the original $SU(3)$
solution, and this can be progressively shifted toward curve \green{B} and into curve
\blue{C} as we change the integration constants. }}
\end{center}
\end{figure}

Due to the complexity of the solutions, the field behaviour is somewhat difficult to
determine by mere inspection. However, one can verify that the symmetry-transformation does
not affect the asymptotic development, and so the same set of states are accessed on the
ellipse at early and late time as in the $SU(3)$ model. However, the intermediate evolution
is substantially, and interestingly, different. In Fig.~\ref{fig:SU3trans2} we plot some
examples on the ellipse.

\begin{figure}[htbp]
\begin{center}
\includegraphics[angle=-90,width=9cm]{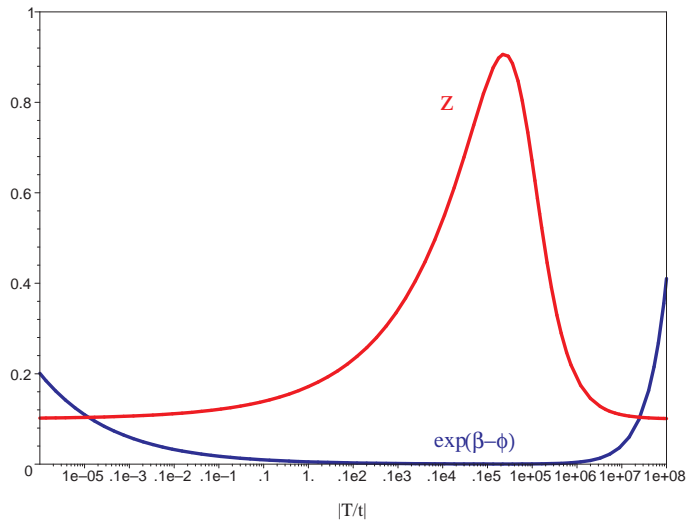}
\caption[$SU(3)$ transformed case 2 brane behaviour]{ \label{fig:SU3trans2brane} \textit{
The plot shows the transformed evolution of the brane \red{$z$}, and the parameters
\blue{$\epsilon_k \propto \exp\left(\beta-\phi\right)$}. Notice that the latter satisfy
$\epsilon_k \ll 1$ while the brane is evolving, and that the brane reverses direction whilst
still strictly inside the orbifold interval. }}
\end{center}
\end{figure}

The particularly interesting feature of these new solutions is the motion of the brane.
Specifically, for certain special choices of constants, the brane can ``bounce'' and
spontaneously reverse direction midway through its evolution. Moreover, a thorough
investigation of the parameter space reveals that it is possible to make $\epsilon_{k}\ll 1$
whilst $z$ is undergoing this bounce strictly inside the orbifold interval. This is shown in
Fig.~\ref{fig:SU3trans2brane}.
%=======================================================================================================FIGURE

To see that this behaviour is indeed a consequence of the solutions in
Eq.~\eqref{eq:su3trans}, one can proceed in the following qualitative fashion. First, we
recognise that the effect of the symmetry transformation is to switch-on the combination
$\dot{\nu}-\chi\dot{z}$ to a non-zero value. This then acts as a driving force that modifies
the original $SU(3)$ evolution of $z$. Secondly, we recognise that this ``modification'', at
a practical level, amounts to additively mixing the $SU(3)$ behaviours of $z$ and $\nu$
together (see Eq.~\eqref{finitetrans}). So not only is $z$ affected, but it is affected by a
non-trivial mixing together with the behaviour of $\nu$. Thirdly, the original $SU(3)$ field
$\nu$ can \emph{already} be made to reverse direction for particular choices of constants
(see Eq.~\eqref{eq:su3nu}). Hence, once mixed, the transformed  $z$ solution \emph{also}
inherits this bouncing behaviour.

As before, this behaviour is subject to $\alpha'$ corrections. However, the strength of
these corrections can always be adjusted such that, when the brane is bouncing, the
corrections are extremely small and so under control. Of course, the corrections cannot be
made arbitrarily small for \emph{all} time, but they can always be made arbitrarily small
over significant periods of time when the brane is moving. To achieve this one simply tunes
$c_{2} \ll 1$, which has the effect of setting $\beta \gg 0$ in the vicinity of the bounce.

These bouncing solutions richly extend the results of the previous sections. We now see that
the effective supergravity action Eq.~\eqref{4daction} admits exact solutions where the
brane evolves in a regime with $\epsilon_{k} \ll 1$, has small $\alpha'$ corrections, is
strictly between the boundaries, and can also reverse direction mid-course. These effects
were not at all obvious from the exactly integrable $SU(2)$ and $SU(3)$ Toda models, and yet
can be generated by judiciously applying symmetries of the equations of motion. We also
reiterate that no explicit potential was required to induce these effects; the reversal is a
natural outcome of the non-linearly coupled cosmology.

%===========================================================================================

\section{Perturbations}

In the previous section we presented several new classes of cosmological solution to
heterotic M-theory, and found that the four-dimensional scale-factor
$a=\exp\left({\alpha}\right)$ always satisfies $a \sim |t-t_{0}|^{1/3}$. Switching to
conformal time $\eta$ defined by $d\eta=a^{-1}dt$, this translates into $a\sim
|\eta-\eta_{0}|^{1/2}$. This means that on the $(+)$ branch we always have an expanding,
decelerating universe, whilst on the $(-)$ branch we always have a contracting, inflationary
universe. This behaviour is to be expected, for the cosmology we are studying has no
explicit potentials and so $a$ remains unaffected by the other fields. We will now consider
in more detail the inflationary epoch and the generation of perturbations on the $(-)$
branch.

As with the familiar PBB scenarios, the inflationary period on the $(-)$ branch is
characterised by a comoving Hubble length $|d(\ln a)/d\eta|^{-1}=2|\eta-\eta_{0}|$ that
decreases as we take $\eta \rightarrow \eta_{0}$ and approach the Big Bang singularity from
below. Consequently, a given comoving scale starting inside the Hubble radius as $\eta
\rightarrow -\infty$ automatically becomes larger than the Hubble radius as $\eta\rightarrow
\eta_{0}$. Therefore, on the $(-)$ branch one can produce super-horizon scale perturbations
merely from kinetic-driven inflation, without the use of any potentials. This is considered
an interesting alternative to conventional inflation on the $(+)$ branch, since PBB
scenarios do not require special choices of potential or slow-roll conditions. Given this,
it is interesting to consider the perturbation spectra of our fields on the $(-)$ branch,
and see whether there any useful scale-invariant modes. Not surprisingly, we will be able to
utilise the techniques developed in PBB cosmology to aid our calculations. We will also see
that the factor of $3$ in the kinetic terms for $\beta$ and $\chi$, which complicated the
classification of the scalar-field manifold (see Ref.~\cite{Copeland:2005mk}) has
interesting consequences for the spectral indices of the fields.

We will begin by considering perturbations around a special background where the axions
$\chi,\sigma,\nu$ have been set to constants, and where the conserved quantity in
Eq.~\eqref{eq:zconserve} has been set to zero. In this case the brane $z$ remains static for
all time, and entirely decouples from the equations of motion of $\beta$ and $\phi$. The
fields $\beta, \phi$ then exhibit standard rolling-radii behaviour with unrestricted
parameters, and no transitions on the ellipse occur. Although this special ``vacuum''
situation does not incorporate any interesting brane displacements in the background, it
proves to be a much simpler situation that can be solved analytically. Later, we will
comment on perturbations around more general backgrounds, including the various Toda models
and their symmetry transforms. In the meantime, we note that in the simple vacuum case the
$\beta$ and $\phi$ perturbations remain coupled to the metric perturbations, and produce
adiabatic perturbations with the same steep $n=4$ blue spectra that occurs in PBB
cosmology~\cite{Lidsey:1999mc,Brustein:1994kn}. In contrast, the fields $z,\sigma,\chi,\nu$
with constant background values are decoupled from the metric perturbations, and produce
isocurvature perturbations $\delta z,\delta \sigma,\delta \chi,\delta \nu$ with different
spectra.

The first-order, gauge-invariant perturbation equations for $\delta z,\delta \sigma,\delta
\chi,\delta \nu$ in conformal time are given by
\begin{align}
\delta z^{\prime\prime} + (2\alpha^\prime+\beta^\prime-\vphi^\prime)\delta z^{\prime} + k^{2}\delta z &= 0 \nonumber \\
\delta \sigma^{\prime\prime} + 2(\alpha^\prime-\vphi^\prime)\delta \sigma^{\prime} +
k^{2}\delta \sigma &=
 -4qz(\beta^\prime-\vphi^\prime)\delta \nu^{\prime} +8qz\chi\beta^\prime\delta z^\prime  \label{eq:sigma_p}\\
\delta \chi^{\prime\prime} + 2(\alpha^\prime-\beta^\prime)\delta \chi^{\prime} + k^{2}\delta \chi &= 0 \nonumber \\
\delta \nu^{\prime\prime} + (2\alpha^\prime-\beta^\prime-\vphi^\prime )\delta\nu^{\prime} +
k^{2}\delta \nu &= -2\chi\beta^\prime\delta z^\prime \label{eq:nu_p}.
\end{align}
Here a $\prime$ denotes a derivative with respect to $\eta$, and $k$ is the comoving
wavenumber of the perturbation. In order to solve for these four isocurvature perturbations
we will use techniques familiar from the PBB literature, with $z$ replacing an axion. This
involves making an appropriate conformal transformation on the metric into each axion's
frame so as to eliminate the coupling to $\beta,\phi$, and then solving the resulting
perturbation equations in the usual manner (see
Refs.~\cite{Lidsey:1999mc,Copeland:1998ie,Copeland:1997ug}). However, before we do this we
need to deal with the awkward source terms on the right-hand sides of Eq.~\eqref{eq:sigma_p}
and Eq.~\eqref{eq:nu_p}. The presence of the bulk-brane field $z$ on the right-hand side of
Eq.~\eqref{eq:sigma_p}, rather than a true axion, slightly complicates the situation as we
cannot simply set $z=0$ as we can with $\chi$ in Eq.~\eqref{eq:nu_p}. Recall that our theory
is only valid when $z\in (0,1)$. Instead, we must deal with the source terms by choosing
appropriate combinations of perturbations: $\delta A = \dsigma + 4qz\dnu$ and $\delta B =
\d\nu - \chi\dz$ \footnote{Unlike $z$, we are always free to set $\chi=0$. As such, we can
choose $\delta B=\d\nu$. }.

Following the calculations of Ref.~\cite{Copeland:1998ie}, we can now define a new metric
for each field's frame by making a conformal transformation on the Einstein metric:
$\bar{g}_{j\;\mn}=\Omega_{j}^{2}g_{\mn}$. Our conformal factors $\Omega_j$ are explicitly
given by
\begin{align}
\Omega_{z}^{2} = e^{\beta-\vphi}  \,,\quad \Omega_{A}^{2} = e^{-2\vphi}      \,,\quad
\Omega_{\chi}^{2} =  e^{-2\beta}  \,,\quad \Omega_{B}^{2} = e^{-\beta-\vphi}
\label{eq:conf_fact}
\end{align}
These conformal transformations lead to a different scale factor $\bar{a}_{j}=\Omega_j a$ in
each frame, depending on each fields coupling to $\beta,\phi$. As we are considering static
axions and bulk brane, $\beta,\phi$ behave as simple rolling-radii fields with fixed
parameters that lie at one point on the ellipse, Eq.~\eqref{eq:ellipse}, for all
time.\footnote{Consequently, we will now drop the subscripts $i,f$ that label the initial
and final rolling-radii powers, as the fields $\beta,\phi$ remain in the same rolling-radii
states for all time. Also, note that we can pick $(p_{\beta},p_{\phi})$ from \emph{anywhere}
on the ellipse, irrespective of the time branch. } Explicitly, in conformal time they
satisfy
\begin{equation}
a=a_*|\eta|^{\frac{1}{2}}   \quad,\quad \beta=\frac{3}{2}p_{\beta} \ln |\eta| + \beta_{0}
\quad,\quad \vphi= \frac{3}{2} p_{\vphi}\ln |\eta| + \vphi_{0}
\end{equation}
where $a_*$ is a constant, and we have conveniently set $\eta_0=0$. Using these background
solutions one can show that
\begin{align}
\bar{a}_{j}=\bar{a}_{*j}|\eta|^{\half+r_j} \quad{\rm with}\quad %
r_{z} = \frac{3}{4} (p_{\beta}-p_{\vphi}) \,,\quad r_{A} = -\frac{3}{2}p_{\vphi} \,,\quad
r_{\chi} = -\frac{3}{2}p_{\beta}  \,,\quad r_{B} = -\frac{3}{4}(p_{\beta}+p_{\vphi})
\label{eq:scale_fact}
\end{align}
where the $\bar{a}_{*j}$ are a set of constants. In these new frames we then find that the
perturbation equations can be recast in the form
\begin{equation}
\delta x_{j}^{\prime\prime}+2\bar{\alpha}_{j}^{\prime}\delta x_{j}^{\prime}+k^{2}\delta
x_{j}= 0 \nonumber
%\label{eq:mod_pert}
\end{equation}
where $\delta x_{j} = (\delta z, \delta A, \delta\chi, \delta B)$ and
$\bar{\alpha}_{j}^{\prime}=\bar{a}_{j}^\prime/\bar{a}_{j}$ is the Hubble rate in each
conformal frame. The solution for our isocurvature perturbations, after normalising at early
time, is then given by (see Ref.~\cite{Copeland:1998ie})
\begin{align*}
\delta x_j =
\kappa\sqrt{\frac{\pi}{m_{j}k}}\exp\left[\frac{i\pi}{4}\left(1+2|r_j|\right)\right]
\frac{(-k\eta)^{1/2}}{\bar{a}_{j}} H_{|r_j|}^{(1)}(-k\eta)
\end{align*}
Here the $m_j$ are given by $m_z=2q$, $m_{A}=4$, $m_{\chi}=12$, $m_{B}=8q$, and $H_J^{(1)}$
is the Hankel function of the first kind and order $J$. Defining the power spectrum
$P_{\delta x}$ and its spectral index $n_{\delta x}$ for a general perturbation $\delta x$
as
\begin{align*}
P_{\delta x}=\frac{k^{3}}{2\pi^2}|\delta x|^{2} \quad {\rm and} \quad n_{\delta x}-1=\frac{d
\ln P_{\delta x}}{d \ln k} \nonumber
\end{align*}
we find the spectral index for each of the isocurvature perturbations is given by
\begin{align*}
n_{\delta x_j}= 4-2|r_j|
\end{align*}

Looking at the definitions of the $|r_j|$ given in Eq.~\eqref{eq:scale_fact}, we see how the
spectral indices are dependent on the coupling of $z$ and the axions to $\phi,\beta$ and
consequently their expansion powers, $p_\phi$ and $p_\beta$. Inserting the specific
couplings for each field and considering the range of background solutions yields
\begin{align*}
&n_{\delta A} = 4-3|p_\vphi| \,:\quad \in [4-2\sqrt{3},4] \quad \sim \quad [0.54,4], \\
&n_{\delta z} = 4-\frac{3}{2}|(p_{\beta}-p_{\vphi})|\,,\quad n_{\delta\chi} = 4-3|p_\beta|
\,,\quad n_{\delta B}=  4-\frac{3}{2}|(p_\beta+p_\vphi)|\,:\quad  \in [2,4]
%\quad{\rm while}\quad.
\end{align*}
Thus we find that our perturbation $\delta A$ has the classic axion perturbation spectrum
familiar from PBB calculations, and can provide a scale-invariant spectrum. In contrast, the
bulk brane and other axion perturbations cannot provide a scale-invariant spectrum, a result
similar to the one obtained in Ref.~\cite{DiMarco:2002eb}.

One can also write the spectral indices as a function of a single variable by using the
ellipse constraint, Eq.~\eqref{eq:ellipse}. This reveals
\begin{align*}
&n_{\delta z} = 4- \left|\pm\sqrt{4-3p_{\vphi}^2} - 3p_{\vphi}\right| \,,\quad
n_{\delta A} = 4-3|p_\vphi|  \,,\quad  \\
&n_{\delta\chi} =  4-\sqrt{4-3p_{\vphi}^2}\,,\quad n_{\delta B}= 4-
\left|\pm\sqrt{4-3p_{\vphi}^2}+ 3p_{\vphi}\right|
\end{align*}
where $p_{\vphi}\in[-2/\sqrt{3} , 2/\sqrt{3}]$ for the vacuum case. One should remember that
the choice of $\pm$ sign must be consistently applied across all the spectral indices, and
that both signs are always valid choices (as we do \emph{not} have to satisfy
$p_{\beta}-p_{\phi}>0$ in the vacuum case).

If one is familiar with PBB calculations the above result may be surprising, as axion
perturbations derived from actions very similar to ours will usually \emph{all} have
spectral indices in the range $[4-2\sqrt{3},4]$. (See, for example, the variety of
dilaton-moduli-axion systems discussed in
Refs.~\cite{Lidsey:1999mc,Copeland:1998ie,Copeland:1997ug,Bridgman:2000kk}). This change is
a direct consequence of the coupling of our fields to $\beta$, which unlike in PBB cosmology
has a factor of $3$ in its kinetic term. This then affects the range of $p_\beta$ through
the ellipse condition. One cannot change this result by rescaling $\beta$'s kinetic term as
this rescales $\beta$'s coupling to the fields and moves the effect into the $r_j$
definitions. This then leaves $\delta A$ as the single perturbation capable of producing a
scale-invariant spectrum.

So far we have only been considering the ``vacuum'' solutions where $z$ and the axions
remain constant. However, generalising these solutions to the case with moving brane and
axions remains an open question, due to the sheer complexity of the solutions considered.
One can begin by truncating off the axions and considering the perturbations of the $\phi,
\beta, z$ $SU(2)$ action of Section \ref{sec:su2}. In this case one can use an
$SL(2,\mathbb{R})$ symmetry of the truncated action~\footnote{This symmetry is not related
to the $SL(2,\mathbb{R})$ generalised $T$-duality we have discussed in this paper, and does
not remain when considering the full, untruncated action.} to solve for the perturbation
$\dz$ around a \emph{moving}-brane $SU(2)$ background, by applying the $SL(2,\mathbb{R})$
symmetry to perturbations $\dphi, \dbeta, \dz$ around a \emph{static}-brane $SU(2)$
background. One then finds that this ``rotated'' $\dz$ isocurvature perturbation retains the
spectrum $n\in [2,4]$~ (see Ref.~\cite{DiMarco:2002eb}). However, the effect of this
rotation on the remaining axionic equations leaves a non-trivial calculation.

As a result, we can only conjecture that a scale-invariant mode persists in perturbations
around the Toda model backgrounds and their symmetry transforms. However, it is certainly
true that all of the solutions we have considered will \emph{asymptotically approach} the
``vacuum'' scenario. Moreover, when one applies the constraints on $p_\vphi$ in the various
classes of solutions one finds that, in a least one of the asymptotic limits, the
rolling-radii regime which leads to $\delta A$ producing a scale-invariant spectrum is
accessible. Hence, we can always generate a scale-invariant mode in one of the asymptotic
limits, even if we cannot determine whether such a mode can also be generated at
intermediate times.

%=============================================CONCLUSION======================================

\section{Conclusion}

We have presented several new classes of cosmological solution to the four-dimensional
effective supergravity description of heterotic M-theory. This theory contain seven fields:
the four-dimensional scale-factor $\alpha$, the modulus $\beta$ measuring the separation of
the orbifold planes, the axion $\chi$ related to the graviphoton field, the dilaton $\phi$
measuring the average Calabi-Yau volume, the axion $\sigma$ related to the bulk three-form,
the field $z$ locating the position of the M5 brane, and the axion $\nu$ representing the
self-dual two-form on the brane worldvolume. To linear-order in the moduli-dependent
parameters $\epsilon_{k}$ $(k=1,2)$, all fields except $\alpha$ can be described by the
following scalar-field Lagrangian
\begin{align*}
\mathcal{L} =
\frac{3}{4}(\partial\beta)^{2}+3e^{-2\beta}(\partial\chi)^{2}+\frac{1}{4}(\partial\phi)^{2}+
\frac{1}{4}e^{-2\phi}\left(\partial\sigma + 4qz\partial\nu\right)^{2}
+\frac{1}{2}qe^{\beta-\phi}(\partial z)^{2}+2qe^{-\beta-\phi}(\partial\nu-\chi
\partial z)^{2}
\end{align*}
We have attempted to identify as many exact solutions to this system as possible, by
identifying special constraints on the fields that simplify the analysis. The only
previously known solution to this Lagrangian, as described in Ref.~\cite{Copeland:2001zp},
is found when $\mathcal{L}$ is consistently truncated to the form
\begin{align*}
\mathcal{L}_{SU(2)} =\frac{3}{4}(\partial\beta)^{2}+
\frac{1}{4}(\partial\phi)^{2}+\frac{1}{2}qe^{\beta-\phi}(\partial z)^{2}
\end{align*}
The fields $\beta,\phi,z$ then form an exactly-solvable $SU(2)$ Toda model, with the brane
$z$ undergoing single displacements. In this paper we have identified three new solutions in
addition to this $SU(2)$ Toda solution. The first new solution was found by consistently
truncating $\mathcal{L}$ to the different form
\begin{align*}
\mathcal{L}_{SU(3)} =
\frac{3}{4}(\partial\beta)^{2}+3e^{-2\beta}(\partial\chi)^{2}+\frac{1}{4}(\partial\phi)^{2}
+\frac{1}{2}qe^{\beta-\phi}(\partial z)^{2}
\end{align*}
using the conditions $\partial\nu-\chi \partial z = \partial\sigma + 4qz\partial\nu = 0$. By
switching off these two terms, one finds that the reduced set of fields $\beta,\chi,\phi,z$
span an integrable $SU(3)$ Toda model, and can be solved for exactly. This $SU(3)$ model
allows for \emph{double} displacements of the brane $z$, and the $SU(3)$ solutions can
always be made reliable with $\epsilon_{k} \ll 1$ over a certain period of time during this
double displacement. However, the brane must leave the compact space in any solution that
has reliable $\epsilon_{k} \ll 1$ regime at some point.

Next, we applied to the $SU(2)$ and $SU(3)$ models the symmetry transformations derived and
discussed in our companion Ref.~\cite{Copeland:2005mk}. This enabled us to derive two new
and distinct cosmological solutions. The properties of these new solutions were then
discussed at length, and it was found that the reliability of the solutions had been
radically affected. This is ultimately due to the $SL(2,\mathbb{R})$ subgroup of symmetries,
which acts as a ``generalised'' set of $T$-duality transformations given by
\begin{align}\label{eq:Tduality1}
T'= \frac{aT-ib}{icT+d} \quad,\quad Z'= \frac{Z}{icT+d} \quad, \quad S'=S-
\frac{icqZ^2}{icT+1}\quad \mathrm{with} \quad  a,b,c,d \in \mathbb{R} \quad, \quad ad-bc=1
\end{align}
This is not ordinary $T$-duality, as all \emph{three} complex fields $T,S,Z$ are affected.
Using these symmetries, it was then found in the $SU(2)$-transformed solutions that the
brane can undergo a single displacement entirely within the orbifold interval with
$\epsilon_{k} \ll 1$ throughout. It was also found in the $SU(3)$-transformed solutions that
the brane field $z$ can undergo two successive displacements of \emph{opposite} sign and so
reverse direction. The specific conditions under which this reversal occurs are as follows.
Firstly, set $\partial \sigma+4qz\partial\nu = 0$ so that the axion $\sigma$ is decoupled
from the other fields. Then the scalar-field lagrangian $\mathcal{L}$ reduces to the simpler
form
\begin{align*}
\mathcal{L'}=\frac{3}{4}(\partial\beta)^{2}+3e^{-2\beta}(\partial\chi)^{2}+
\frac{1}{4}(\partial\phi)^{2}+ \frac{1}{2}qe^{\beta-\phi}(\partial
z)^{2}+2qe^{-\beta-\phi}(\partial\nu-\chi
\partial z)^{2}
\end{align*}
One then proceeds by setting $\partial\nu-\chi\partial z =0$ and solving the system as an
$SU(3)$ Toda model, but then \emph{restoring} the $\partial\nu-\chi\partial z$ term to a
general, non-zero value by applying an $SL(2,\mathbb{R})$ symmetry. In particular, the
fields $z,\nu$ transform as a doublet under $SL(2,\mathbb{R})$, and so we can solve the
system with general $\partial\nu-\chi\partial z$ by ``rotating'' from a solution where it is
zero. As a consequence, the equations of motion arising from the reduced lagrangian
$\mathcal{L'}$ have been completely solved in this paper. Further, by tuning the sign and
magnitude of the $-\chi\partial z$ contribution generated by the symmetry application, one
can modify the overall velocity of the brane so that it comes to rest and reverses
direction. This is a particularly interesting feature arising from the coupling with $\chi$,
whose presence in the kinetic term $\partial \nu-\chi\partial z$ is due to the need for a
gauge-covariant derivative in five dimensions. This reversing behaviour, which we have
occasionally called a ``bouncing'' solution, can also be made to occur entirely within the
orbifold interval with $\epsilon_{k} \ll 1$ throughout.

%Consequently, it is likely that this reversing feature persists even in the full system
%with $\partial \sigma+4qz\partial\nu \neq 0$, and is likely to appear in other theories
%where there is an

As such, all of the transformed solutions demonstrate a rich new variety of M5 brane
behaviours, and new, trustworthy regions of solution space emerge that had not previously
been identified. In particular, we conjecture that reversing brane solutions will exist in
other corners of string theory beyond heterotic M-theory. One can pin down reasonably clear
``minimum conditions'' for this reversal to occur, as follows. Firstly, at least one modulus
should be active, such as the dilaton $\phi$, whose coupling to the brane kinetic term will
induce the brane $z$ to undergo a single displacement. Secondly, there should also be an
active combination proportional to a cross-coupling between an axion field and $\partial z$.
This second combination can then be adjusted so that the brane turns around at some point
during its motion.

As an interesting corollary, we then considered the isocurvature perturbation spectra
produced by the model in an inflationary contracting (PBB) phase. In the ``vacuum'' case we
found that one of the isocurvature modes -- the one associated with the axions $\sigma$ and
$\nu$ -- is able to produce a scale-invariant spectrum. Furthermore, we found that
\emph{all} of the solutions considered will asymptotically approach this vacuum case in at
least one asymptotic limit, and so a scale-invariant perturbation spectrum can always be
generated asymptotically when perturbing around \emph{any} of the solutions we have studied.
However, the detailed structure of the perturbation spectrum at intermediate times has not
yet been computed in its full generality, and it would be interesting to study this problem
in greater depth, perhaps in a manner analogous to the numerical approach developed in
Refs.~\cite{Cartier:2003jz,Cartier:2004zn}.

Finally, we note that the methodologies employed in this paper have much wider
applicability. For example, the Toda model solution method, as extensively detailed in
Refs.~\cite{Kostant:1979qu, Lukas:1996iq}, is not restricted to scalar-field systems arising
from heterotic M-theory, and could be readily utilised in other areas of string theory.
Likewise, it is equally plausible that other braneworld K\"ahler metrics may possess useful
symmetry groups, which can be used to transform subsystems of fields into new patterns of
behaviour. In light of this, it would be interesting to clarify the origin of the special
$SL(2,\mathbb{R})$ symmetry group that we have found, and understand the general conditions
under which reversing brane behaviour occurs in string and M-theory.

%%%%%%%%%%%%%%%%%%%%%%%%%%%%%%%%%%%%%%%%%%%%%%%%%%%%%

\bibliographystyle{plain}

\section*{Appendix A}

In this Appendix we present certain additional details of the $SU(3)$ Toda model derivation.
We do this because the derivation is rather complicated, particularly the manner in which
one must change time-gauges and judiciously redefine constants.

To begin with, we know that the vectors $\mathbf{q}_{1},\mathbf{q}_{2}$ are proportional to
the two simple root-vectors of $SU(3)$. Utilising this fact, we can choose a basis for the
space $(\alpha,\beta,\phi)$ that is adapted to the underlying $SU(3)$ symmetry, and so
consists of vectors $\mathbf{e}_{0},\mathbf{e}_{1},\mathbf{e}_{2}$ satisfying
\begin{align}
<\mathbf{e}_{0},\mathbf{e}_{0}>=-1 \quad, \quad %
<\mathbf{e}_{0},\mathbf{e}_{1}>\:=\:<\mathbf{e}_{0},\mathbf{e}_{2}> \:=\:0 \quad, \quad %
\mathbf{e}_{1} =\frac{3}{8}\mathbf{q}_{1} \quad, \quad %
\mathbf{e}_{2} =\frac{3}{8}\mathbf{q}_{2}
\end{align}
A choice of basis compatible with these conditions is given by
\begin{align}
\mathbf{e}_{0} =(\sqrt{3},0,0) \quad,\quad \mathbf{e}_{1}= \frac{3}{8}(0,-1,1) \quad,\quad
\mathbf{e}_{2} = \frac{3}{8} (0,2,0)
\end{align}
We now write the covariant vector $G\mbox{\boldmath $\alpha$}$ as the following sum
\begin{align}
G\mbox{\boldmath$\alpha$} = \sum_{i=0}^{2}\rho_{i}(\tau)\mathbf{e}_{i}
\end{align}
and insert this time-dependent expansion into the equations of motion to find the evolution
of the ``modes'' $\rho_{i}$. Choosing the convenient gauge $n=3\alpha$ (or $E=1$) one finds
\begin{align}
\ddot{\rho}_{0}=0\\
\ddot{\rho}_{1} +\frac{4}{3}u_{1}^{2}e^{2\rho_{1}-\rho_{2}} = 0\\
\ddot{\rho}_{2} +\frac{4}{3}u_{2}^{2}e^{2\rho_{2}-\rho_{1}} = 0\\
-\dot{\rho}_{0}^{2} +
\frac{3}{4}\left(\dot{\rho}_{1}^{2}-\dot{\rho}_{1}\dot{\rho}_{2}+\dot{\rho}_{2}^{2}\right)
+2U = 0
\end{align}
The general solution to these equations is now easy to come by, and takes the form
\begin{align}
\rho_{0}&=-k_{0}(\tau-\tau_{0})\\
\rho_{1}&= -\ln{g_{1}(\tau)}\\
\rho_{2}&= -\ln{g_{2}(\tau)}
\end{align}
where $k_{0},\tau_{0}$ are constants. The functions $g_{1},g_{2}$ are given by a sum over
the collection of weight vectors
\mbox{$\mbox{\boldmath$\Lambda$}_{1}=\{(0,-1),(-1,1),(1,0)\}$}, %
\mbox{$\mbox{\boldmath$\Lambda$}_{2}=\{(-1,0),(1,-1),(0,1)\}$} of the fundamental
$\mathbf{3}$ and $\bar{\mathbf{3}}$ representations of $SU(3)$. Concretely, if we define the
matrix of vectors
\begin{align}
\mbox{\boldmath$\lambda$}_{ij}= \left(\begin{array}{c}
                                       \mbox{\boldmath$\Lambda$}_{1}\\
                                       \mbox{\boldmath$\Lambda$}_{2}
                                        \end{array} \right)
                              =\left(\begin{array}{ccc}
                                   (0,-1)& (-1,1) & (1,0)   \\
                                   (-1,0)&(1,-1) &(0,1)
                              \end{array} \right)
\end{align}
then for $i=1,2$ the functions $g_{i}$ are given by
\begin{align}
g_{i} =\sum_{j=1}^{3} %
a_{ij} \: \mathrm{exp}\left[\mbox{\boldmath$\lambda$}_{ij}\cdot
\left(\mathbf{k}\tau-\mbox{\boldmath$\tau$}\right)\right]
\end{align}
where the positive constants $a_{ij}$ are
\begin{align*}
a_{11} &= \frac{4u_{2}^2}{3}\left(\frac{2k_{1}-k_{2}}{P}\right)& %
a_{21} &= \frac{4u_{2}^2}{3}\left(\frac{2k_{2}-k_{1}}{P}\right)\\
a_{12} &= \frac{4u_{1}^2}{3}\left(\frac{k_{1}+k_{2}}{P}\right)& %
a_{22} &= \frac{4u_{2}^2}{3}\left(\frac{k_{1}+k_{2}}{P}\right)&\\
a_{13} &= \frac{4u_{1}^2}{3}\left(\frac{2k_{2}-k_{1}}{P}\right)& %
a_{23} &= \frac{4u_{1}^2}{3}\left(\frac{2k_{1}-k_{2}}{P}\right)&\\
\end{align*}
and $P=(2k_{1}-k_{2})(2k_{2}-k_{1})(k_{1}+k_{2})$. The constant vector
$\mbox{\boldmath$\tau$} =(\tau_{1},\tau_{2})$ is a set of arbitrary time-shifts. The
constant vector $\mathbf{k}=(k_{1},k_{2})$ is restricted to the open Weyl chamber, which
means it is forced to have positive scalar product with the two simple root-vectors as
follows
\begin{align}
(k_{1},k_{2})\cdot(2,-1) &= 2k_{1}-k_{2} >0  \nonumber\\
(k_{1},k_{2})\cdot(-1,2) &= 2k_{2}-k_{1} >0 \;. \label{eq:Weyl_ch}
\end{align}
These two conditions guarantee that $g_{1},g_{2}>0$ so that the logarithms in
$\rho_{1},\rho_{2}$ are always well-defined. Lastly, we must also impose the Friedmann
constraint
\begin{align}
 -k_{0}^{2} +
\frac{3}{4}\left(k_{1}^{2}-k_{1}k_{2}+k_{2}^{2}\right)= 0 \;. \label{eq:cons_toda}
\end{align}

Using the information above, one can arrive at an explicit solution for the fields
$\alpha,\beta,\phi$ and the two additional fields $z,\chi$ that were integrated out. Recall
that
\begin{align}
\mbox{\boldmath$\alpha$} = \left(\begin{array}{c}
                                \alpha\\
                                \beta\\
                                \phi
                                \end{array}\right)
= \sum_{i=0}^{2}\rho_{i}G^{-1}\mathbf{e}_{i} =  \left(\begin{array}{c}
                                                    -\frac{1}{\sqrt{3}}\rho_{0}\\
                                                    \rho_{2}-\frac{1}{2}\rho_{1}\\
                                                     \frac{3}{2}\rho_{1}
                                                    \end{array}\right)
\end{align}
and that
\begin{align}\label{gfuncs}
g_{1}(\tau)&= a_{11}e^{-k_{2}\tau+\tau_{2}}
+a_{12}e^{(k_{2}-k_{1})\tau-(\tau_{2}-\tau_{1})}+a_{13}e^{k_{1}\tau-\tau_{1}}  \nonumber\\
g_{2}(\tau)&=a_{21}e^{-k_{1}\tau+\tau_{1}}+a_{22}e^{(k_{1}-k_{2})\tau-(\tau_{1}-\tau_{2})}+
a_{23}e^{k_{2}\tau-\tau_{2}}
\end{align}
Then we find
\begin{align}
\alpha &= \frac{1}{\sqrt{3}}\:k_{0}\left(\tau-\tau_{0}\right)  \\
\beta &= \frac{1}{2}\ln\left[a_{11}e^{-k_{2}\tau+\tau_{2}}
+a_{12}e^{(k_{2}-k_{1})\tau-(\tau_{2}-\tau_{1})}+a_{13}e^{k_{1}\tau-\tau_{1}}  \right] \nonumber\\
&\quad
-\ln\left[a_{21}e^{-k_{1}\tau+\tau_{1}}+a_{22}e^{(k_{1}-k_{2})\tau-(\tau_{1}-\tau_{2})}+
a_{23}e^{k_{2}\tau-\tau_{2}} \right]  \nonumber \\
\phi &= -\frac{3}{2}\ln\left[a_{11}e^{-k_{2}\tau+\tau_{2}}
+a_{12}e^{(k_{2}-k_{1})\tau-(\tau_{2}-\tau_{1})}+a_{13}e^{k_{1}\tau-\tau_{1}} \right] \nonumber\\
z-z_{0} &= C_{z}\frac{\dot{g_{1}}}{g_{1}}= C_{z} \left[ \frac{
-k_{2}a_{11}e^{-k_{2}\tau+\tau_{2}}+(k_{2}-k_{1})a_{12}e^{(k_{2}-k_{1})\tau-(\tau_{2}-\tau_{1})}
+ k_{1}a_{13}e^{k_{1}\tau-\tau_{1}} }{ a_{11}e^{-k_{2}\tau+\tau_{2}} +
a_{12}e^{(k_{2}-k_{1})\tau-(\tau_{2}-\tau_{1})}+
a_{13}e^{k_{1}\tau-\tau_{1}} } \right]  \nonumber\\
\chi-\chi_{0} &=  C_{\chi}\frac{\dot{g_{2}}}{g_{2}}=
C_{\chi}\left[\frac{-k_{1}a_{21}e^{-k_{1}\tau+\tau_{1}}
+(k_{1}-k_{2})a_{22}e^{(k_{1}-k_{2})\tau-(\tau_{1}-\tau_{2})}
k_{2}a_{23}e^{k_{2}\tau-\tau_{2}} }{a_{21}e^{-k_{1}\tau+\tau_{1}}  +
a_{22}e^{(k_{1}-k_{2})\tau-(\tau_{1}-\tau_{2})} +a_{23}e^{k_{2}\tau-\tau_{2}} } \right]
\nonumber \label{eq:toda_sols}
\end{align}
where $z_{0},\chi_{0}$ are constants of integration, and
\begin{align*}
C_{z}= \left(\frac{9}{8qu_{1}^{2}}\right)^{\frac{1}{2}} \quad, \quad %
C_{\chi}= \left(\frac{3}{16u_{2}^{2}}\right)^{\frac{1}{2}}
\end{align*}
We can then deduce the forms of $\sigma,\nu$ that are compatible with the ancillary
conditions Eq.~\eqref{eq:todatrunc}
\begin{align*}
\nu-\nu_{0}&= \chi_{0}(z-z_{0}) +
C_{z}C_{\chi}\left[\frac{k_{2}(2k_{1}-k_{2})a_{13}e^{k_{1}\tau-\tau_{1}}
+k_{1}(2k_{2}-k_{1})a_{23}e^{-k_{2}\tau+\tau_{2}}}{
a_{11}e^{-k_{2}\tau+\tau_{2}}+a_{12}e^{(k_{2}-k_{1})\tau-(\tau_{2}-\tau_{1})}
+a_{13}e^{k_{1}\tau-\tau_{1}}}\right] \\
\sigma-\sigma_{0}&= -2q\left[2z_{0}\nu + \chi_{0}(z-z_{0})^{2} +\int
\left[(z-z_{0})^{2}\right]^{\centerdot}(\chi-\chi_{0})d\tau \right]
\end{align*}
where $\nu_{0},\sigma_{0}$ are two further constants of integration. Notice that the
integral in the $\sigma$ solution is not elementary, and so cannot be written as a finite
(potentially nested) sequence of logs, exponentials and rational functions of $\tau$.
However, it can sometimes be analytically integrated for fixed choices of the arbitrary
constants, and in any event has a sensible definite integral between fixed $\tau$ limits. In
particular, one can compute the $\sigma$ behaviour numerically for any given set of starting
conditions.

It is useful to understand the asymptotic limits, in order to transform these solutions to
the proper time-gauge $n=0$. One finds that
\begin{align*}
e^{\beta} &\sim e^{\left(2k_{1}-k_{2}\right)\tau/2} \quad \mathrm{and} \quad
e^{\phi}   \sim e^{3k_{2}\tau/2}                    \quad \mathrm{as} \quad \tau \rightarrow -\infty\\
e^{\beta} &\sim e^{\left(k_{1}-2k_{2}\right)\tau/2} \quad \mathrm{and} \quad e^{\phi}   \sim
e^{-3k_{1}\tau/2}                   \quad \mathrm{as} \quad \tau \rightarrow +\infty
\end{align*}

Since $2k_{1}-k_{2}>0$ and $k_{1}-2k_{2}<0$ we see that the orbifold radius $\beta$ always
goes from a state of expansion at early time to a state of contraction at late time. The
same is true of the  modulus $\phi$ that measures the orbifold-averaged Calabi-Yau volume.
These two fields will then have some complicated intermediate transition(s) that smoothly
link these extreme limits. On the other hand, the fields $z$ and $\chi$ always asymptote to
constants in the limits, although these constants are generally different. They too will
undergo some intermediate ``displacement" consistent with the different constant field
values at early and late time.

We now change from logarithmic time $\tau$ to proper time $t$. Since the logarithmic-time
gauge is given by $n=3\alpha$, we can find the relation to proper time by integrating the
defining relation
\begin{align}
dt \equiv e^{n(\tau)}d\tau = e^{3\alpha(\tau)}d\tau = e^{\sqrt{3}k_{0}(\tau-\tau_{0})}d\tau
\end{align}
This gives
\begin{align}
\alpha = \frac{1}{3}\ln\left[\sqrt{3}k_{0}(t-t_{0})\right] \equiv
\frac{1}{3}\ln\left|\frac{t-t_{0}}{T_{0}}\right|
\end{align}
where $t_{0}$ is a finite integration constant. This leads to two disconnected time branches
corresponding to the choices $t-t_{0}>0, T_{0}>0$ and  $t-t_{0}<0, T_{0}<0$, both of which
lead to a well-defined positive argument for the logarithm. The regime $t-t_{0}>0,T_{0}>0$
will be referred to as the ``positive-time'' or simply $(+)$ branch, whilst the sector
$t-t_{0}<0, T_{0}<0$ will be dubbed the ``negative-time'' or simply $(-)$ branch. The
physics in the time interval $\tau \in (-\infty,+\infty)$ is mapped to these two regions in
the following way. The early time $\tau \rightarrow -\infty$ regime with expanding $\beta,
\phi$ corresponds to $t-t_{0}\rightarrow 0$ on the $(+)$ branch and $t-t_{0}\rightarrow
-\infty$ on the $(-)$ branch, while the late time regime $\tau \rightarrow \infty$
corresponds to $t-t_{0}\rightarrow \infty $ on the $(+)$ branch and $t-t_{0}\rightarrow 0$
on the $(-)$ branch. It should be noted these two time branches in $t$ are physically
separated by an unavoidable curvature singularity at $t=t_{0}$, despite the fact that in the
$\tau$-gauge we had only one physical region. After this gauge change, typical terms in
$\rho_{1},\rho_{2}$ will then scale as
\begin{align*}
\left|\frac{t-t_{0}}{T_{0}}\right|^{\lambda_{i}'} \quad, \quad
\left|\frac{t-t_{0}}{T_{0}}\right|^{-\lambda_{i}'}
\end{align*}
respectively, where $\lambda_{i}'=(k'_{1},k_{2}'-k_{1}',-k_{2}')$ and
\begin{align*}
k_{1}' = k_{1}T_{0} \quad, \quad  k_{2}' = k_{2}T_{0}
\end{align*}

Note that $k_{1}',k_{2}'$ can be of either sign depending upon the sign of $T_{0}$ and so
the choice of branch. On the \emph{positive} branch we have $k_{1}',k_{2}'>0$ so that
$k_{1}'=|k_{1}'|, k_{2}'=|k_{2}'|$.  In this gauge the fields $\beta,\phi$ scale
asymptotically at early-time as
\begin{align*}
\beta \sim p_{\beta,i}\ln|t-t_{0}| \quad, \quad \phi \sim p_{\phi,i}\ln|t-t_{0}|
\end{align*}
where we have defined the two constants
\begin{align*}
p_{\beta,i} = \frac{1}{2}\left(2|k_{1}'|-|k_{2}'|\right) \quad, \quad p_{\phi,i} =
\frac{3}{2}|k_{2}'|
\end{align*}
The Friedmann constraint Eq.~\eqref{eq:cons_toda} then reduces to the familiar ellipse
condition
\begin{align*}
p_{\phi,i}^{2}+3p_{\beta,i}^{2} = \frac{4}{3}
\end{align*}
Moreover, the Weyl chamber constraints  $2|k_{1}'|-|k_{2}'|>0,\: 2|k_{2}'|-|k_{1}'|>0$
translate into $p_{\beta,i}>0, \: \delta<0$. So the positive branch is associated to
$\delta<0$ with the further additional constraint that $\delta_{\beta}=-2p_{\beta,i}<0$.
Conversely, on the \emph{negative} time branch we find the opposite results, since we follow
the ellipse trajectories backward. Hence, we find that $\delta,\delta_{\beta}>0$.

Since all the fields involved are scalars, we are now free to substitute $\tau$ in terms of
$t$ in all the solutions. These will then be the $n=0$ forms for the solutions. To make
these solutions look ``nice'', however, one must carefully redefine certain constants. If
one defines new timescales $T,T_{\beta}$ via
\begin{align*}
\left|\frac{T}{T_{0}}\right|^{\delta} =
\frac{a_{12}}{a_{11}}\exp\left[-(k_{1}-2k_{2})\tau_{0}+(\tau_{1}-2\tau_{2})\right]
\quad,\quad \left|\frac{T_{\beta}}{T_{0}}\right|^{\delta_{\beta}} =
\frac{a_{22}}{a_{21}}\exp\left[(2k_{1}-k_{2})\tau_{0}-(2\tau_{1}-\tau_{2})\right]
\end{align*}

then the resulting solutions look relatively simple. They can be made even simpler by
defining the ubiquitous fractional combinations
\begin{align*}
\theta_{z} = \frac{a_{13}a_{21}}{a_{12}a_{22}} \quad,\quad \theta_{\chi} =
\frac{a_{23}a_{11}}{a_{12}a_{22}}
\end{align*}

Finally, we have decided to write the solution for $\beta$ so that it scales with respect to
$T$ at early time rather than the natural choice $T_{\beta}$, and so the ratio $T/T_{\beta}$
has been absorbed into the additive offset $\beta_{0}$. This brings the $\beta$ solution
closer to the old $SU(2)$ form and simplifies the vector notation, but at the expense of
introducing the ratio $T/T_{\beta}$ into the constraints.

\end{document}